\documentclass[structabstract]{aa}
\usepackage{graphicx}
\usepackage{natbib}
\usepackage{lscape}
\def\ebv{\mbox{$E_{B-V}$}}
\def\halpha{\mbox{H$\alpha$}}
\def\hbeta{\mbox{H$\beta$}}
\def\hdelta{\mbox{H$\delta$}}

\def\lesssim{\mathrel{\hbox{\rlap{\hbox{\lower4pt\hbox{$\sim$}}}\hbox{$<$}}}}
\def\gtrsim{\mathrel{\hbox{\rlap{\hbox{\lower4pt\hbox{$\sim$}}}\hbox{$>$}}}}

\usepackage{txfonts}
\begin{document}
\title{IFU observations of the GRB~980425/SN~1998bw host galaxy:
  emission line ratios in GRB regions \thanks{Based on observations collected at the European Southern Observatory, Chile: programme ID 077.D-0488(A), and archive programmes: 64.H-0375(A), 66.D-0576(A) and 275.D-5039(A).} }

\titlerunning{IFU observations of the GRB~980425 host galaxy}
   \author{L. Christensen    \inst{1}
          \and P.~M. Vreeswijk  \inst{2}
          \and J. Sollerman  \inst{2,3}
          \and C.~C. Th\"one    \inst{2}
          \and E. Le Floc'h  \inst{4}
          \and K. Wiersema   \inst{5}
   }

%   \offprints{L. Christensen}  
   \institute{European Southern Observatory, Casilla 19001, Santiago 19, Chile,
     \email{lichrist@eso.org}
    \and  Dark Cosmology Centre, Niels Bohr Institute, University of
   Copenhagen, Juliane Maries Vej 30, 2100 {\O}, Denmark
    \and Department of Astronomy, AlbaNova, 106 91, Stockholm University, Sweden 
    \and Spitzer Fellow, Institute for Astronomy, 2680 Woodlawn Drive, Honolulu HI 96822, USA
    \and University of Leicester,  University Road, Leicester, LE1 7RH, United Kingdom 
}
   \date{Received 3 April 2008/Accepted 16 July 2008}
   
%%%%%%%%%%%%%%%%%%%%%%%%%%%%%%%%%%%%%%%%%%%%%%%%%%%%%%%%%%%%%%%%%%%%%%%%
     
   \abstract
%Context (optional) 
{The collapsar model predicts that the progenitors of Gamma-ray Bursts
  (GRBs) are metal poor in Fe group elements. The existence of low
  metallicity stellar populations could manifest itself in the
  characteristics of the GRB site immediate environment in the host
  galaxy. }
%Aims
{We analyse the strong emission lines from the sub-luminous host
  galaxy of GRB~980425, which showed the first connection with a
  supernova explosion (SN~1998bw). The host is of sufficient size to
  allow detailed resolved spectroscopy of individual \ion{H}{ii}
  regions and to search for regions with peculiar properties close to
  the the GRB site.}
%method
{Using integral field spectroscopy with VIMOS we study most of the
  high surface brightness part of the host including the \ion{H}{ii}
  region where the supernova and GRB occurred.}
%results
{The star formation rate, reddening, equivalent width and stellar mass
  in the GRB region is similar to other \ion{H}{ii} regions in the
  host. Extreme values arise in the only region that shows emission
  lines from Wolf-Rayet stars, a region that is located 800 pc in
  projection from the GRB site. Strong emission line diagnostics of
  all \ion{H}{ii} regions imply oxygen abundances between 0.3 and 0.8
  solar with the lowest values arising in the WR and GRB regions.
  Including uncertainties from the metallicity diagnostics, all
  metallicities are similar to within 3$\sigma$. We demonstrate that
  there is a good agreement between the luminosity weighted and mass
  weighted specific star formation rates (SSFR) in individual young
  \ion{H}{ii} regions.  While the global average of the SSFR is
  similar to high redshift GRB hosts, there are significant variations
  between individual resolved \ion{H}{ii} regions.  Comparing the
  measured emission line ratios of low redshift GRB hosts to
  theoretical models and observations of field galaxies, we find that
  GRBs are present in different environment metallicities while the
  regions of their origin are consistently very young. Similar line
  ratios of GRB hosts compared with those of the WR region can arise
  in spatially unresolved galaxies with bright \ion{H}{ii} regions
  close to the GRB location. }
%Conclusions (optional)
{}

\maketitle
%

%
%________________________________________________________________

\section{Introduction}
The association of the Gamma-ray Burst \object{GRB~980425} with the
supernova \object{SN~1998bw} \citep{gal98} was initially questioned
because of the low redshift of the supernova host galaxy \object{ESO
  184-G82}, $z=0.0085$.  The corresponding isotropic energy output of
GRB~980425 was low (E~$\sim~10^{49}$ erg) compared with other GRBs
known at that time (at $z~\gtrsim~1$ with E$~\sim~10^{52}$ erg). Since
then, more apparently under-luminous, low redshift GRB events have
been detected. The definite proof of a supernova connection originated
with the discovery of SN Type Ic signatures in the afterglow spectra
of a couple of long duration GRBs
\citep{hjorth03,stanek03,malesani04}.  While the energetics of
GRB~980425 suggested an underluminous event, there was no indication
that its accompanying supernova was peculiar for a GRB-SN
\citep{kaneko07}.

About 500 GRBs have been observed and optical afterglows have been
detected for about a third\footnote{\tt
  http://www.mpe.mpg.de/$\sim$jcg/grbgen.html}.  While bursts detected
by the BeppoSAX satellite typically had redshifts $z\sim1$, those
detected by the Swift satellite \citep{gehrels04} have an average
redshift of 2.2 \citep{jakobsson06}\footnote{\tt
  http://www.astro.ku.dk/$\sim$pallja/GRBsample.html}.  The small
angular size of the high redshift host galaxies (typically less than a
few arcseconds) implies that a closer examination of the nature of the
hosts requires integrated spectra or broad band images for the
faintest hosts. Images of GRB hosts indicate that GRBs are located
preferentially in sub-luminous \citep{lefloch03} actively
star-forming, young galaxies \citep{christensen04b}. The distribution
of GRB afterglow locations within their hosts follow in general the
bright continuum light distribution of the host galaxies
\citep{bloom02a,fruchter06}.

Models of single star collapsar progenitors of long duration bursts
predict that they are more likely to reside in metal poor environments
\citep{woosley93}.  Low metallicities in the surrounding medium of
GRBs are in general measured by absorption line studies of GRB
afterglows at $z\gtrsim2$
\citep{savaglio03,vreeswijk04,chen05,dessauges06,prochaska07}. This
technique is used mostly to infer iron group elements along the line
of sight towards the GRB. High resolution spectra of GRB afterglows
have indicated that the medium probed by the spectra resides at
distances larger than several tens of pc from the burst site
\citep{prochaska06b,chen07}, and a detailed study of one burst
(GRB~060418) discovered that the environment probed by the absorption
lines was at a distance of 1.7~kpc from the burst site
\citep{vreeswijk07}.  The GRB afterglow does therefore not probe the
medium immediate surrounding the burst itself, but rather the
interstellar medium of the host; this implies that perhaps we are
unable to estimate the metallicity of the progenitor directly using
this technique.

Since the hosts of long duration bursts are undoubtedly star-forming
galaxies, the emission lines from the integrated optical spectra can
be used to infer the metallicities provided that the redshift of the
GRB is less than approximately one where the strong lines are visible
in the optical. This technique is generally used to derive the oxygen
abundance. Diagnostics of strong emission lines for the hosts of
\object{GRB~980425}, \object{GRB~020912}, \object{GRB~030329},
\object{GRB~031203}, and \object{GRB 060218} have indicated sub solar
metallicities
\citep{prochaska04,hammer06,gorosabel05,sollerman05,wiersema07}, and
systematically lower host galaxy metallicities than found in nearby
ordinary Type Ic supernovae sites \citep{modjaz07}.  It is, however,
unclear how measurements derived from the observed integrated GRB host
spectra relate to the overall host properties, or to the site in which
the GRB exploded.  If the host galaxy is clearly extended, an analysis
of the spatially resolved emission lines can uncover the
characteristics of the precise GRB region, or alternatively the
specific environment conditions required to make a GRB.
\citet{thoene08} investigated the spectra of the host of GRB~060505,
and found that the GRB site was the youngest and most metal poor of
all regions probed by the slit spectrum.

For the host of GRB~980425/SN~1998bw (\object{ESO~184-G82}), a dwarf
SBc galaxy, the SN occurred in a star-forming region in the spiral
arm. The region was identified by high spatial resolution images
acquired with HST/STIS \citep{fynbo00} and confirmed by the
observation of a fading source in this region \citep{sollerman02}.
The location of the SN was offset by a projected distance of 0.8~kpc
from the strongest star-forming region in the host, which had distinct
spectral signatures of Wolf-Rayet stars \citep{hammer06}. The spectrum
of the SN region did not have any detected WR features.  Other GRB
hosts at $z\lesssim0.1$ did not show clear signatures of WR stars in
emission either \citep{margutti07,wiersema07}.  At higher redshifts,
the absence of \ion{C}{iv} $\lambda$ 1550 absorption lines in winds of
WR stars can be explained by the afterglow, which ionises the
surrounding medium to a large radius (\citet{chen07}, but see also
\citet{starling05})

All previous analyses of emission lines from GRB hosts were limited to
regions within the slit. Hence, the data could be affected by
slit-losses and leading to an under-estimated integrated star
formation rate (SFR).  In this paper, we present data for the
GRB~980425 host galaxy obtained with integral field spectroscopy.  The
data from the VIMOS integral field unit (IFU) presented in
Sect.~\ref{sect:data} allows us to study a significant fraction of the
surface of the host with a resolution of 0.27~kpc, and to analyse most
of its high surface brightness regions (see Fig.~\ref{fig:hostim}).
We use the data to compile maps of emission lines and their ratios and
determine the reddening, SFRs, specific SFRs, metallicities,
densities, temperatures, and kinematic and stellar masses, as
presented in Sect.~\ref{sect:results}. We compare the emission line
ratios with those of other GRB hosts in Sect.~\ref{sect:ratios}, and
show evidence that, while abundances may vary, the regions are in all
cases very young.  In Sect.~\ref{sect:disc}, we discuss the
implications for the unresolved high redshift GRB hosts. We adopt a
cosmology with H$_0$=70 km~s$^{-1}$~Mpc$^{-1}$ throughout the paper.

%=================================
\section{Observations and data reduction}
\label{sect:data}

\begin{figure*}[t!]
\centering
\resizebox{12cm}{!}{\includegraphics[bb=0 0 520 264, clip]{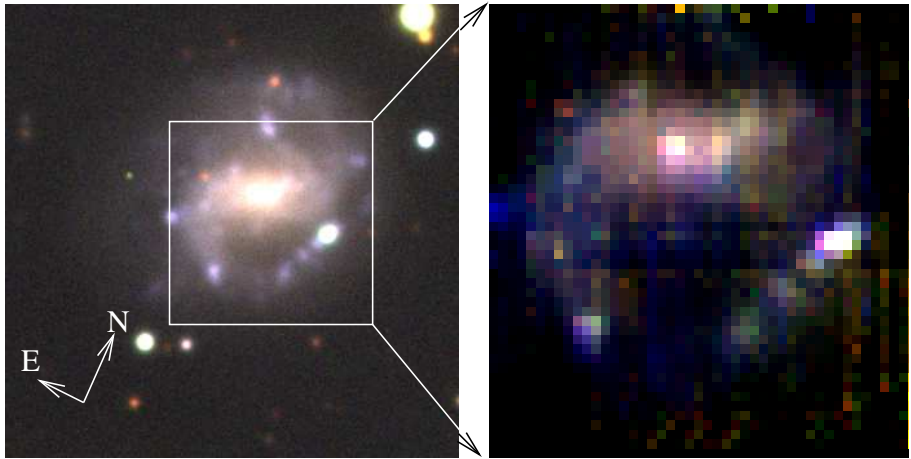}}
\caption{Colour composite images of the host galaxy from archive FORS
  $BVR$ broad band images (left) and broad band images extracted from
  the VIMOS data cube (right).  Each of the pixels in the VIMOS image
  has a corresponding spectrum. The WR region is the brightest and
  bluest one to the West of the centre, and the SN region is seen as a
  fainter extension adjoining this region to the South.  The size of
  the VIMOS image is 30\arcsec\ on a side, corresponding to 5.3 kpc at
  the redshift of the galaxy.  {\it See the electronic edition of the
    Journal for a colour version of this figure}. }
\label{fig:hostim}
\end{figure*}

The host galaxy was observed on six different clear or photometric
nights in April and May 2006 in service mode at the Very Large
Telescope UT3 {\it Melipal} with the VIMOS integral field
spectroscopic mode. The high spectral resolution setup of VIMOS
consists of four quadrants of 400 spectra each, recorded on 4
CCDs. The field of view (FOV) when using the 0\farcs67 lens array is
27\arcsec$\times$27\arcsec. The size of the host as measured in FORS
images obtained from the ESO archive (64.H-0375 and 66.D-0576(A), PI:
F. Patat), is about 1\arcmin\ in diameter, while the high surface
brightness emission is roughly half that size. We obtained a total
integration time of five hours in three different settings as listed
in Table~\ref{tab:obslog}. Since the host galaxy emission occupies the
entire IFU FOV (see Fig.~\ref{fig:hostim}), separate 200~s
observations of the sky offset by 2 arcmin were used for sky
background subtraction.  The wavelength coverage of the VIMOS data
from the three settings allowed us to cover all of the strong emission
lines used for abundance determinations. The low spectral resolution
data also covered the [\ion{O}{ii}]~$\lambda$3727 line. At the
redshift of the host, this emission line is detected at the observed
wavelength of 3759~${\AA}$, where the transmission of the VIMOS IFU is
about 10 times lower than its maximum value.

The data were reduced with the reduction package P3d \citep{becker02}
and the following steps. After bias subtraction, the location of the
1600 spectra and the traces were identified using continuum lamp
spectra obtained immediately after the science observations. Arc line
spectra taken after science exposures were extracted. The science
spectra were extracted, wavelength calibrated, and flat fielded using
the (wavelength dependent) transmission function determined from the
continuum spectra. The calibrated science data were arranged in a 3D
data cube. Each data cube was corrected for atmospheric
extinction. The sky background was subtracted using separately reduced
sky background data cubes.

The 4 quadrants of the combined data cubes were flux calibrated
separately using observations of spectrophotometric stars placed in
each of the four quadrants.  Spatial shifts between the individual
data cubes were determined and the cubes were combined, using routines
similar to standard image combination procedures in IRAF, while
masking out the dead spectra that were located mainly at the edge of
the FOV.  We ignored the differential atmospheric refraction effects,
since the observations were mostly taken at air masses lower than 1.4.
This should theoretically produce a maximum shift of $\sim$1 spaxel (a
spaxel `spatial picture element', is defined to be a single spectrum
in the array) between the blue and the red end of the spectral range
\citep{filippenko82}.  However, the maximum spatial shift in the data
cubes was measured to be less than 0.2 pixels by a cross correlation
technique. Since we used dithered pointings that were offset by
3\arcsec, the resulting FOV in the reduced and combined data cube was
30\arcsec$\times$30\arcsec.

\begin{figure*}[t!]
\centering
\resizebox{14cm}{!}{\includegraphics{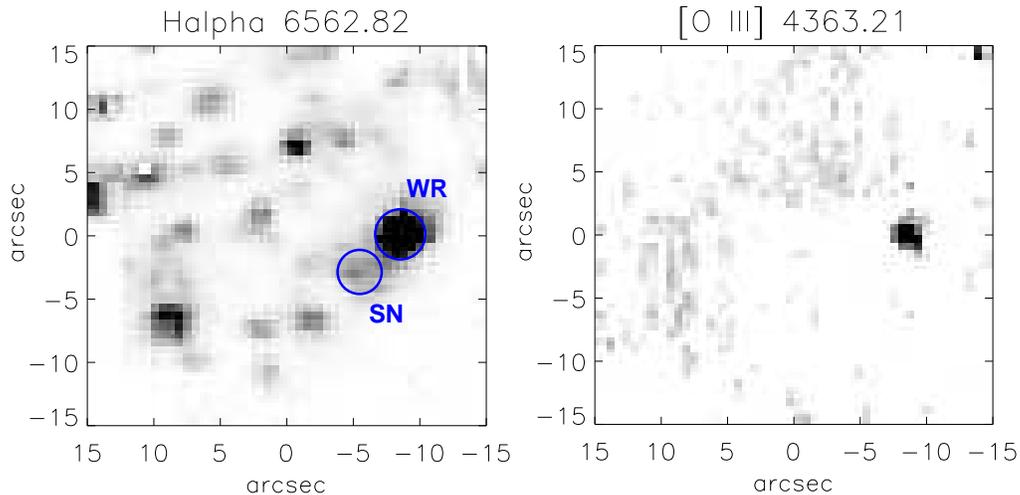}}
\caption{Maps of a typical strong emission line (left: \halpha), and a
  faint one (right: [\ion{O}{iii}]~$\lambda$4363). The region that is
  bright in [\ion{O}{iii}] is the only region where emission lines
  from WR stars are detected. The circles have a radius of 2\farcs5
  corresponding to 440 pc.}
\label{fig:maps}
\end{figure*}

To check the absolute flux calibration, we re-reduced broad band FORS
$B$, $V$, and $R$ images from the ESO archive. We then extracted the
spectra for different regions from the VIMOS high spectral resolution
cubes, and convolved each spectrum with the Bessell filter
transmission functions to calculate the broad band magnitudes
\citep{bessell90}.  For individual regions, such as the SN and WR
region and the entire VIMOS FOV in Fig.~\ref{fig:hostim}, the
magnitudes and colours agree with those for the FORS images to within
10\% accuracy.  It is impossible to apply a simple scaling to the
VIMOS data cube to improve the absolute spectrophotometric calibration
for all regions within the FOV. The sensitivity of the VIMOS-IFU is
known to show variations depending on the exact location of the
spectrophotometric standard star \citep[see][]{monreal06}, which
affects the absolute flux calibration. Finally, the data cube was
corrected for the foreground Galactic reddening of \ebv=0.059 mag
\citep{schlegel98}.

Even though the VIMOS data were taken in good seeing conditions, using
the 0\farcs67 spatial sampling provided an effective resolution
element of 1\farcs5, or 0.27 kpc at the redshift of the galaxy.

\begin{table*}
\centering
\begin{tabular}{lclccc}
\hline
\hline
\noalign{\smallskip}
Date & Integration & Setting & $\lambda$ & $R$ & Seeing \\
     & time (s)    &         & ($\AA$)   & ($\lambda/\Delta\lambda$)    &  ($\arcsec$)\\
\noalign{\smallskip}
\hline
\noalign{\smallskip}
2006 April 25 & 3$\times$600 & LR blue   & 3600--6800 & 250  & 0.9\\
2006 May   24 & 2$\times$600 & LR blue   & 3600--6800 & 250  & 0.9\\
2006 April 29 & 3$\times$1800& HR orange & 5045--7475 & 2700 & 0.7\\
2006 May 3  & 2$\times$1800 & HR orange  & 5045--7475 & 2700 & 0.8\\
2006 May 24 & 2$\times$1800 & HR blue    & 3970--6230 & 2600 & 1.0\\
2006 May 25 & 1$\times$1800 & HR blue    & 3970--6230 & 2600 & 0.4\\
\noalign{\smallskip}
\hline
\end{tabular}
\caption{Log of the observations. $R$ gives the average spectral
  resolution as measured from the data. The resolution varies by 10\%
  from one spectrum to the next. The seeing is the one recorded by the
  observatory DIMM monitor, and not the effective spatial resolution
  obtained from the data cube.}
\label{tab:obslog}
\end{table*}

%--------------------------------------------
\section{Results}
\label{sect:results}
\subsection{Emission line maps and spectra}
IFU data cubes allow us to estimate fluxes both from images and
spectra. Two-dimensional images are created by coadding monochromatic
slices (equivalent to channel maps in the radio astronomy community)
in the data cube along the wavelength direction. These can be analysed
with standard imaging methods. One-dimensional spectra are created by
selecting some pixels or `spaxels' in the data cube and co-adding each
wavelength separately. To extract one dimensional spectra and
visualize the data cubes, we used the program QFitsView created by
Thomas Ott.

To estimate line fluxes, we used two methods: fitting of lines in
spectra with Gaussian profiles and a simple integration of narrow-band
images created from the data cube.  Since the emission lines from the
different regions are not strictly Gaussian in shape, we cannot rely
on a fully automated line-fitting routine to extract line fluxes. To
create emission line maps and derive fluxes for different regions, we
use the second method, integrating over 8 slices from the data cube in
the spectral direction centred on the measured wavelength
$(1+z)\lambda_{\mathrm{lab}}$.  This corresponds to twice the size of
the spectral \textit{FWHM} and is equivalent to integrating over a
wavelength interval between 4.4 and 4.7 ${\AA}$ about the lines. The
emission lines are barely resolved even with the high spectral
resolution ($\sim$110 km~s$^{-1}$ at 5000 {\AA}), and no emission
lines in any of the regions have widths significantly larger than the
spectral resolution. The emission line fluxes determined from the two
methods are consistent to within 10\% accuracy.

The continuum flux was determined by fitting a first order polynomial
over a region about the emission line in each of the individual
spectra.  The continuum region was selected free of other emission
lines and examined interactively. The backgrounds were subtracted to
provide the pure emission line maps. Two examples are presented in
Fig.~\ref{fig:maps}, which shows the \halpha\ emission line map (left)
and [\ion{O}{iii}]~$\lambda$4363 (right). Only the brightest
\ion{H}{ii} region in the host shows detectable emission from the
latter, faint, temperature-sensitive line. This region also shows
emission lines with characteristic Wolf-Rayet features
\citep{hammer06}. We investigated the emission line maps to look for
WR emission lines (He or N lines) in other regions, but these lines
are below the detection limit in the data cube. Following
\citet{hammer06}, the brightest region is denoted `WR', while the
star-forming region in which the SN/GRB occurred is denoted `SN'. The
SN occurred in an \ion{H}{ii} region that is located 5 arcsec to the
South-East of the centre of the WR region, or 800 pc in projection at
the redshift of the host. In the spectrum extracted from the WR
region, we identify 45 different lines within the spectral
coverage\footnote{The spectrum presented in \citet{hammer06} also has
  significantly more lines than listed in the paper.}.  In other
regions, the number of detectable lines is significantly smaller. This
paper considers mostly strong emission lines, since these can be
identified throughout the host galaxy, allowing a fair comparison
between different regions.

\subsection{Emission line fluxes}
\label{sect:emline}
Table~\ref{tab:emlines} lists the line fluxes and derived properties
of several distinct and spatially resolved \ion{H}{ii} regions
identified in the \halpha\ map. Most regions listed in the table are
labeled with their coordinates relative to the \halpha\ map. A
significant fraction (34\%) of the integrated emission line flux
within the VIMOS FOV originates in the WR region. The final row in the
Table presents the sample mean and standard deviation as a measure of
the spread of values in the different \ion{H}{ii} regions.

The \halpha\ and \hbeta\ Balmer emission line fluxes have been
corrected for an underlying stellar absorption component with the
equivalent width EW(\halpha)$\approx$EW(\hbeta)~=~2.6~{\AA}, which is
an appropriate value for young stellar populations
\citep{gonzalezdelgado99}. We observe no evidence for an absorption
line about the \hbeta\ line in any individual spaxel. While the
absorption line should be present in any stellar population, the
non-detection in individual spectra is because the continuum emission
is faint and the signal-to-noise ratio in each spaxel is typically
only a few. In the combined galaxy spectrum an underlying stellar
absorption component is evident with an equivalent width of
$\sim$2~{\AA}.

The equivalent width is defined as \(\mathrm{EW}=\int_{\lambda}
(f_{\mathrm{line,\lambda}}/f_{\mathrm{cont,\lambda}}-1)d\lambda\),
where $f_{\mathrm{line,\lambda}}$ is the measured emission line flux
and $f_{\mathrm{cont,\lambda}}$ is the continuum flux.  We correct the
integrated line flux for an absorption of 2.6 {\AA} such that
\(f_{\mathrm{line,cor}}=f_{\mathrm{line}}+2.6 \times
f_{\mathrm{cont}}\).  For regions with fainter Balmer emission line
fluxes and lower equivalent widths, the correction to the \hbeta\ line
can be as substantial as 15\%, relative to a typical correction of a
few percent for the brightest lines. This implies that the extinction
will be over estimated and the metallicity under estimated, if this
correction is not applied.  For the brightest emission line regions
the corrections are small.

The [\ion{O}{ii}] emission line in the low spectral resolution data
has considerably larger uncertainty than the other bright emission
lines. Furthermore, no spectrophotometric standard star was observed
with the low spectral resolution setup at the dates of the
observations. Instead, we used standard star observations acquired at
times that differed by about one week, and relied on the fact that the
overall shape of the transmission function of VIMOS is stable with
time. To derive the emission line fluxes, we compared the extracted
\hbeta\ fluxes for the \ion{H}{ii} regions and applied a scaling to
the [\ion{O}{ii}] emission line.

If we compare the line fluxes of the WR and SN regions with those
reported for long slit spectra in the literature
\citep{sollerman05,hammer06}, as well as FORS spectra obtained from
the ESO archive (Program ID 275.D-5039(A), PI: Patat), we must take
into account the different spatial regions sampled.  Due to different
effects, such as seeing-dependent slit losses, an uncertain placement
of the slit, different spectral resolutions, seeing, as well as
reduction procedures to extract the one dimensional spectra, it is
difficult to compare directly our extracted fluxes with values in the
literature \citep[e.g.][]{sollerman05,hammer06}. As an example, the
fluxes from the literature agree with our results obtained from the
archive FORS long slit spectrum only to within a factor of
approximately 2. For these reasons, we cannot compare the VIMOS
spectra with those derived from long slit spectra. While slit losses
affect the long slit spectra, the ratios of the emission line fluxes
should remain the same as in the IFU data provided that the same
region is extracted. The \halpha/\hbeta\ emission line ratio is used
frequently to estimate the extinction (see Sect.~\ref{sect:extinc})
and in the SN region this ratio is used to infer both zero extinction
\citep{sollerman05} and $A_V$=1.02 \citep{hammer06}, while we find
$A_V$=1.95 in the IFU data, similar to $A_v$=1.73 in
\citet{savaglio08}.  For other emission line ratios of lines with
similar wavelengths, we find differences of up to a factor of
two. This could be due to spectrally blended lines or blending of
separate \ion{H}{ii} regions due to seeing.

The broad band magnitudes derived for these regions agree with those
for the FORS imaging data to within 5--10\%, which we take as
representative of the uncertainty in the emission line fluxes.  This
uncertainty is significantly higher than the pure statistical
uncertainty derived from the number counts extracted for each line and
region in the non-flux calibrated data ($<3$\% for the faintest lines
in Table~\ref{tab:emlines}). The uncertainties in the \halpha\ and
\hbeta\ equivalents width are 14\% since they are derived from a ratio
of two quantities each with an uncertainty of 10\%. The uncertainties
of the reddening and abundances are derived by propagating errors. The
uncertainties of all reddening values are 0.12 mag.  Propagating the
flux errors for the abundances determination (see
Sect.~\ref{sect:metal}), the uncertainty is only 0.03 for the oxygen
abundances. Since the scatter in the abundance diagnostics used to
derive the oxygen abundance itself is 0.14 dex \citep{pettini04}, this
will dominate the overall uncertainty. 

Similarly, taking into account a wide range of relevant temperatures
to derive the N abundance we find a 0.2 dex uncertainty in the N/O
ratio.

For the oxygen abundance derived in Sect~\ref{sect:metal}, we use the
solar value 12+log(O/H)=8.66 \citep{asplund04}.

\begin{landscape}
\begin{table}
\extracolsep{60pt}
%\vspace*{2cm}
\begin{tabular}{l@{} r@{.}l @{}r@{.}l  @{}r@{.}l  @{}r@{.}l  r@{.}l  r@{.}l
    r@{.}l  r@{.}l @{}r@{.}l @{}cccc @{}c @{}c@{}}
\hline
\hline
\noalign{\smallskip}
Region & \multicolumn{2}{c}{[\ion{O}{ii}]} &
\multicolumn{2}{c}{\hbeta} & \multicolumn{2}{c}{[\ion{O}{iii}]} &
\multicolumn{2}{c}{\halpha} & \multicolumn{2}{c}{[\ion{N}{ii}]} &
\multicolumn{2}{c}{[\ion{S}{ii}]} & \multicolumn{2}{c}{[\ion{S}{ii}]}
& \multicolumn{2}{c}{EW(\halpha)} & \multicolumn{2}{c}{EW(\hbeta)} 
& $B$ & $V$ & $R$  & \ebv &
12+log(O/H) & log(N/O) \\
 &   \multicolumn{2}{c}{$\lambda$3727} & \multicolumn{2}{c}{} &
\multicolumn{2}{c}{$\lambda$5007} & \multicolumn{2}{c}{}  &
\multicolumn{2}{c}{$\lambda$6584} & \multicolumn{2}{c}{$\lambda$6716}
& \multicolumn{2}{c}{$\lambda$6731} & \multicolumn{2}{c}{$({\AA})$} & 
\multicolumn{2}{c}{$({\AA})$} &  (mag)  & (mag) & (mag) & (mag) \\
%(1) & \multicolumn{2}{c}{(2)} & \multicolumn{2}{c}{(3)} & \multicolumn{2}{c}{(4)} & \multicolumn{2}{c}{(5)} & \multicolumn{2}{c}{(6)} & \multicolumn{2}{c}{(7)} & \multicolumn{2}{c}{(8)} & (9) & (10) & (11) &(12) & (13) \\  
\noalign{\smallskip}
\hline
\noalign{\smallskip}
Galaxy & 2400&0 & 445&0 & 1293&0 & 1793&0  & 197&0 & 283&0   & 68&0 &
67& & 13& &  15.31 & 15.08  & 14.85 & 0.30 & 8.27 & --1.50 \\
WR reg & 694&6 & 174&0 & 815&0 & 603&0 & 47&9 & 51&9 & 39&78 &744& &
141& & 18.64 & 
18.27  & 18.3 & 0.17 & 8.16 & --1.50 \\
SN reg & 26&07 & 7&55 & 18&27 & 44&94 & 5&03 & 9&19 & 6&84 & 178& & 
16& & 20.28 & 
20.21 & 19.8 & 0.63 & 8.30  & --1.28 \\
\hline
\noalign{\smallskip}

WR(--8.7, 0.0) & 436&5 & 142&0 & 719&0 & 419&0 & 30&2 & 30&5 & 24&1 &
1140& & 226& & 
19.16 & 18.72 & 18.92 & 0.03  & 8.14 & --1.42 \\ %r27

SN(--5.3, --2.7)& 8&11 & 2&83 & 7&14 & 19&42 & 2&33 & 4&34 & 3&17 &
208& & 22& & 
21.28 & 21.24 & 21.17 & 0.75 & 8.31  & --1.17 \\ %r26

--11.4, 11.3    & 0&80 & 0&74 & 0&57 & 9&41& 1&13 & 1&51 & 1&13 & 79&
& 6&2 & 
20.96 & 21.12 & 20.74 & 1.29 & 8.47  & --0.64   \\ %r20(41,39)

--8.0, 12.0     & 2&77 & 1&47 & 1&20 & 9&23 & 1&59  & 2&43  & 1&90 &
75& & 12& & 
21.93 & 21.20 & 20.72 & 0.68 & 8.51  & --0.83 \\ %r18(35,39)

--6.7, 8.0      & 1&68 & 1&11 & 1&40 & 7&75 & 0&96 & 1&36 & 1&15 & 38&
&  5&2 & 
21.04 & 20.57 & 20.24 & 0.78 & 8.41  & --0.88 \\ %r24(33,33)

--4.0, 7.3      & 6&30 & 3&43 & 3&89 & 16&08 & 2&01 & 2&50 & 1&83 &
62& & 12& & 
20.59 & 20.22 & 19.92 & 0.43 & 8.41  & --0.98 \\ %r14(29,32)

--2.0, 14.0     & 6&43 & 2&01 & 2&60 & 11&13 & 1&57 & 1&89 & 1&54 &
74& & 12& & 
21.28 & 20.80 & 20.50 & 0.57 & 8.42  & --1.16 \\%r16(26,42)

--1.3, --6.1    & 3&80 & 1&70 & 2&37 & 17&68 & 2&05 & 3&13 & 2&19 &
420& &  16& & 
21.47& 21.52 & 21.37  & 1.12 & 8.38  & --1.00 \\ %r2(25,12)

--1.3, 10.0     & 2&13 & 1&45 & 0&92 & 7&87 & 1&15 & 1&53 & 1&25 & 34&
&  5&6 &
20.67& 20.35 & 20.06 & 0.56 & 8.53   & --0.81 \\ %r21(26,37)

--0.7, 1.3      & 34&1 & 11&80 & 25&51 & 48&51 & 6&29 & 5&04 & 3&61 &
119& &  25& & 
20.14 & 19.65 & 19.38 & 0.32 & 8.34  & --1.15 \\ %r13(24,32)

--0.0, 3.3      & 2&19 & 1&32 & 3&13 & 7&56 & 1&22 & 1&62 & 1&22 & 29&
&  4&5 &
20.65 & 20.21 & 19.94 & 0.60 & 8.36  & --0.84 \\%r22(24,27)

2.0, 12.7       & 0&00 & 0&84 & 0&83 & 5&25 & 0&73 & 1&27 & 1&08  &
230& & 4&9 & 
21.17 & 20.76 & 20.43 & 0.67 & 8.46  & ... \\ %%r17(21,41)

2.7, 1.3        & 4&34 & 4&20 & 3&52 & 17&25 & 2&56 & 2&77 & 1&83 &
296& & 15& & 
20.44 & 20.44 & 20.26 & 0.32 & 8.49  & --0.65 \\ %r12(19,23)

2.7, --4.0      & 6&34 & 0&92& 1&27  & 10&97 & 1&59 & 2&76 & 1&96 &
2411& &  13& & 
21.73 & 21.87 & 21.82 & 1.23 & 8.42  & --1.32 \\ %r3(19,11)&  

5.4, 6.0        & 2&84 & 3&04 & 2&76 & 13&14 & 1&91 & 2&34 & 1&72 &
38& &  7&3 & 
20.23 & 19.86 & 19.62 & 0.36 & 8.48  & --0.62 \\ %r10(15,30)

6.1, 10.7       & 5&59 & 4&57& 4&74  & 16&04 & 2&44 & 3&50 & 2&48 &
58& & 12& & 
 20.25 & 19.98 & 19.85 & 0.18 & 8.46 & --0.70 \\ %r11(14,37)

8.7, 0.6        & 8&15 & 6&65 & 3&40 & 15&95 & 2&04 & 4&53  & 2&84 &
49& &   17& & 
20.16 & 19.90 & 19.37 & 0.00 & 8.54  & --0.83 \\ %r6(10,22)

9.4, --3.3      & 28&3 & 17&39 & 37&42 & 42&96 & 4&77 & 7&03 & 5&20 &
254& &  56& & 
21.24 & 21.02 & 21.07 & 0.00 & 8.32  & --1.00 \\ %r5(9,12) 

10.1, 8.0       & 0&24 & 2&70 & 4&96 & 10&82 & 1&33 & 1&83  & 1&31 &
47& &  8&2 & 
20.26 &  20.18 &  20.19 & 0.29 & 8.35& --0.64 \\%r9(8,33) 

11.4, 4.6       & 21&7 & 12&18 & 27&19 & 27&10 & 4&52 & 4&07& 3&02 &
149& &  19& &
19.63 & 19.43 &  18.94  & 0.00 & 8.37& --0.92 \\ %r8(6,28) 

12.1, 13.3      & 0&00 & 0&77& 0&76 & 3&68 & 0&52 & 0&70 & 0&59 & 38&
&  4&7 & 
21.27 & 20.97 & 20.99  & 0.45 & 8.46 & ...\\  %%r19(6,42) 

14.8, 3.3       & 13&6 & 3&42& 6&00 & 9&25 & 1&00 & 1&92 & 1&34 & 61&
& 15& & 
20.70 & 20.55 & 20.08 & 0.00 &  8.34 & --1.36\\ %r7(1,26) 

14.8, 10.0      & 7&49 & 3&48 & 4&85 & 7&28 & 1&16 & 1&53 & 1&39 & 79&
& 25& & 
21.11 & 21.08 & 20.98 & 0.00 & 8.43  & --1.03\\ %r15(1,36)

\hline
mean     &29&$\pm$94 & 10&$\pm$29 & 38&$\pm$149 & 33&$\pm$85 &3&$\pm$6
&4&$\pm$6 &3&$\pm$5 & 152&$\pm$243 &   24&$\pm$46 & 20.5$\pm$0.6 & 20.3$\pm$0.7&
20.0$\pm$0.6 & 0.46$\pm$0.40  & 8.41$\pm$0.09 & --0.91$\pm$0.34\\
  \noalign{\smallskip}
\hline
\end{tabular}
%\end{center}
\caption{Fluxes for the strong emission lines in units of 10$^{-16}$
  erg~s$^{-1}$~cm$^{-2}$ corrected for Galactic extinction. The
  magnitudes are given in AB units. The first three rows correspond
  \newline to different aperture sizes.  For the WR and SN region the
  radial apertures are 2\arcsec\ defined to include all the flux from
  the regions seen in the \halpha\ image.  Below the first three rows
  \newline the fluxes are extracted from the images in a square
  aperture of $3\times3$ spaxels equivalent to 4 arcsec$^2$ around the
  coordinates in Col. 1.  The magnitudes are calculated from
  \newline the spectra extracted within this aperture. Note that the
  properties vary with the chosen aperture; especially the equivalent
  widths show differences as described in Sect.~\ref{sect:ew}.
  \newline The oxygen abundance and N/O abundance ratio in the two
  last columns are discussed in Sect.~\ref{sect:metal}. The
  uncertainties for line fluxes and magnitudes are $\approx$10\%,
  $<$4\% \newline for the \halpha\ and \hbeta\ EW, 0.12 mag for the
  reddening, and 0.14 dex for the oxygen abundance calibration. The
  uncertainty of the N/O ratio is 0.2 dex. The last row gives the
  \newline  mean of the individual regions, and the uncertainty is
  the standard deviation.}
\label{tab:emlines}
\end{table}
\end{landscape}

\subsection{Kinematics}
\begin{figure}
\resizebox{\hsize}{!}{\includegraphics[bb=48 360 412
    622,clip]{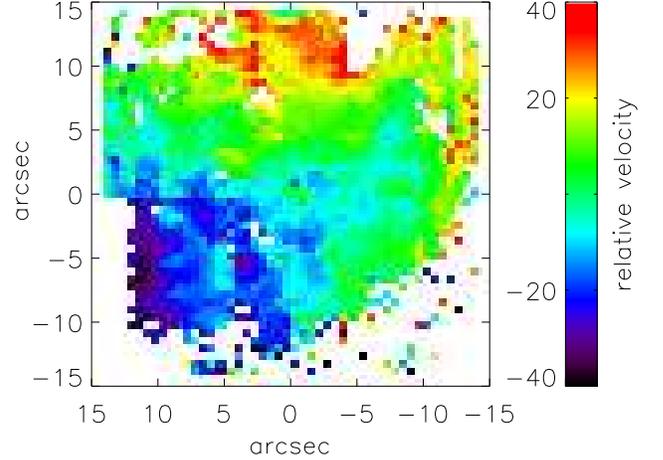}}
\caption{Velocity map derived from fits to the \halpha\ emission line
  in all spaxels. The values are in units of km~s$^{-1}$. {\it See the
    electronic edition of the Journal for a colour version}.}
\label{fig:velo}
\end{figure}

Although the lines are not exactly Gaussian in shape, we use an
automated line-fitting routine to estimate the peak position of the
\halpha\ line and derive the velocity structure of the galaxy
presented in Fig.~\ref{fig:velo}. The zero velocity is defined for the
geometric centre of the galaxy.  The uncertainty in the wavelength
calibration for each spaxel is less than 0.2 {\AA}, which corresponds
to a velocity uncertainty of 10~km~s$^{-1}$.

To derive the kinematic mass of the entire galaxy, it is necessary to
study the rotation curve out to galactocentric radii at which it
flattens, which correspond to the outskirts of the disk. In the
velocity map, there is a gradient that increases all the way to the
edge, so the estimated kinematic mass is limited to within
15\arcsec\ or correspondingly 2.6 kpc in radius.  We model the
rotation of the galaxy by assuming simple rotation of a solid
disk. This is appropriate at least for the central part of the galaxy
because of the presence of a bar in the centre. The only free
parameters are the centre of the disk, the position angle, and the
velocity. The model velocity field is subtracted from the
observations, and the model with the smallest velocity residuals are
taken to present the best fit.  A best-fit model of the velocity field
within the VIMOS FOV provides a maximum rotation velocity of
25~km~s$^{-1}$ at this distance. With these values, the mass is
4$\times$10$^8/(\sin~i)^2$~M$_{\odot}$, where $i$ is the inclination
of the galaxy. For $i\approx50^{\circ}$ estimated from surface
photometry of the high surface brightness emission in the FORS $B$
band image, the lower limit to the galaxy mass is
6$\times$10$^8$~M$_{\odot}$ within the VIMOS FOV. The largest
uncertainty originates from the estimate of the inclination: most of
the galaxy does not have a high surface brightness, and this could
produce an underestimation of the mass by a factor of $\sim$5. The
total galaxy mass derived from a $K$ band image (1.1$\times10^9$
M$_{\odot}$) corresponds to a stellar mass of about twice the value
within the VIMOS FOV \citep{castro-ceron08}.  If the inclination axis
is lower than 50$^{\circ}$, the ratio of the total mass indicated from
the velocity to the stellar mass increases correspondingly, i.e. the
dark matter halo mass increases.

The estimated mass is consistent with the mass-luminosity-metallicity
relation for SDSS galaxies \citep{tremonti04}, which predicts that
12+log(O/H)=8.5$\pm$0.2 for a galaxy of this (stellar) mass. The
dynamical mass estimate, derived above, includes both stellar and dark
matter: a measurement of the stellar mass only should correspond to a
lower expected abundance. In Sect~\ref{sect:metal}, we measure an
oxygen abundance of 12+log(O/H)=8.3, which corresponds to a relatively
small mass galaxy (See also Table~\ref{tab:emlines}).

\subsection{Equivalent widths}
\label{sect:ew}
\begin{figure*}
\centering
\begin{minipage}[c]{0.48\textwidth}
\resizebox{\hsize}{!}{\includegraphics[bb=48 360 432 622,clip]{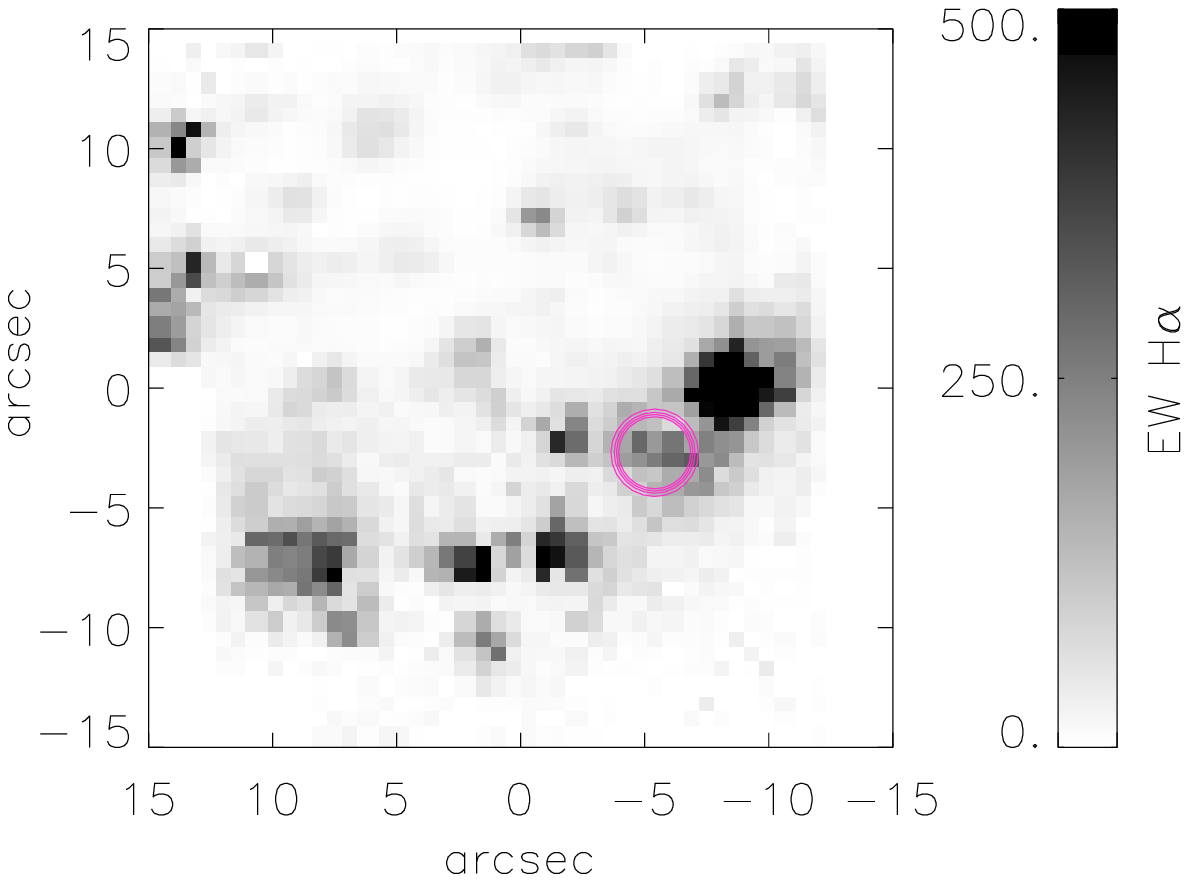}}
\end{minipage}%
\begin{minipage}[c]{0.48\textwidth}
\resizebox{\hsize}{!}{\includegraphics[bb=48 360 432 622,clip]{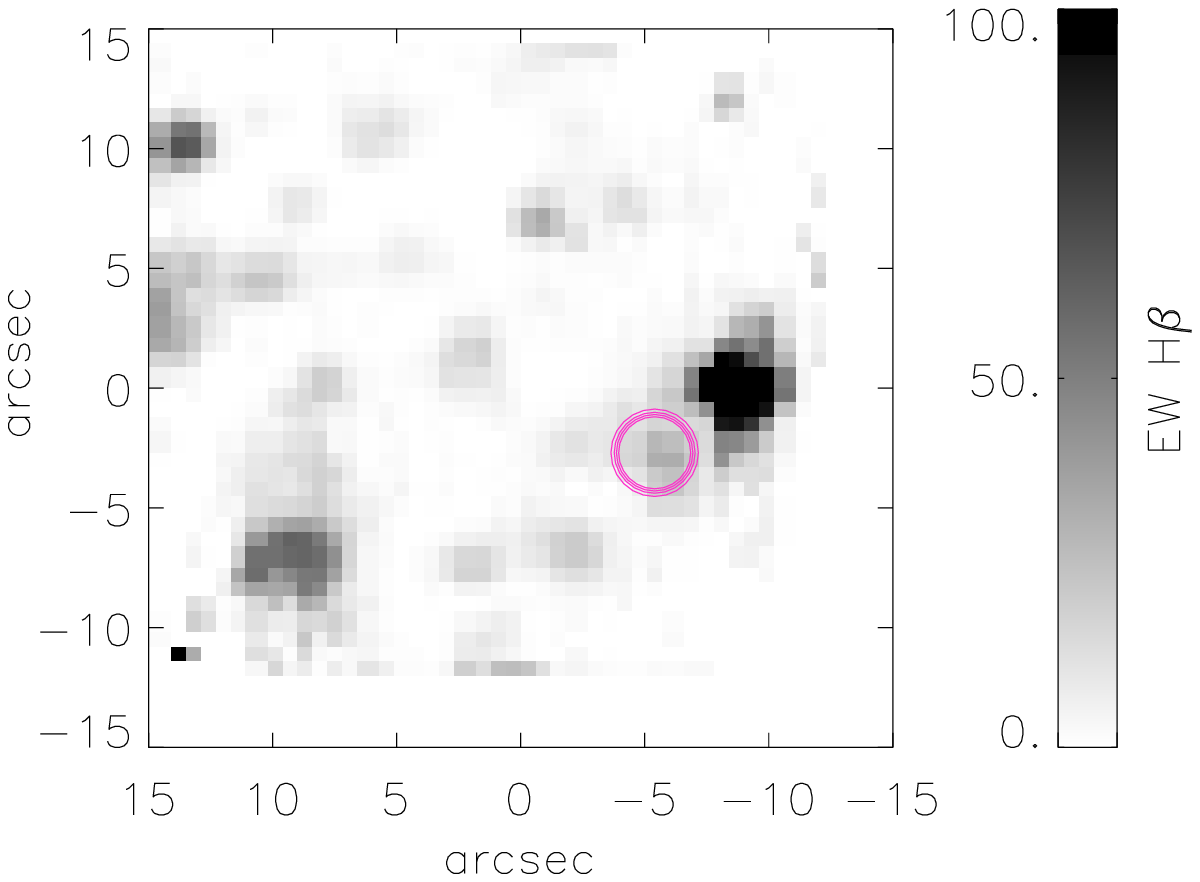}}
\end{minipage}
\caption{Map of the \halpha\ EW (left panel) and map of the \hbeta\ EW
  (right panel). The noise pattern in the lower middle part of the
  \halpha\ EW map is caused by noise in the continuum flux estimate in
  the VIMOS data. In this region, the surface brightness is low (see
  Fig.~\ref{fig:hostim}).  Both panels show that the WR region has the
  largest EW over the face of the galaxy and indicates that this is
  the youngest star-forming region. The circle marks the position of
  the SN.}
\label{fig:ew}
\end{figure*}

The continuum around the \halpha\ and \hbeta\ lines was calculated to
determine the equivalent widths (EWs) corrected for the underlying
stellar absorption. The resulting maps are shown in
Fig.~\ref{fig:ew}. The WR region is distinctive in this image.  In the
WR region, the \halpha\ and \hbeta\ EWs are 744~{\AA} and 141~{\AA},
respectively, within a 2~arcsec radial aperture, while in the SN
region the EWs are 178~{\AA} and 16~{\AA}, respectively. The
\halpha\ value for the SN region is higher by a factor of two than the
measurement of \citet{hammer06}; this disagreement is probably caused
by aperture effects in long slit spectra or differences in the size of
the extracted regions. The fewer the number of extracted spaxels the
larger the EWs. As an example, the EW in the brightest spaxel in the
WR region has an $\halpha$ EW of 1600~{\AA}. According to
instantaneous burst models from Starburst99 \citep{leitherer99}, this
corresponds to an age of 3.1~Myr. Similarly, the brightest spaxel of
the SN region has an $\halpha$ EW of 320~{\AA}, corresponding to an
age of 5.0~Myr. If we analyse the EWs for the other \ion{H}{ii}
regions in the host, we find that the SN region is more representative
of the entire host galaxy, while the WR region is the youngest.

\subsection{Intrinsic reddening}
\label{sect:extinc}
\begin{figure}
\resizebox{\hsize}{!}{\includegraphics[bb=48 360 412 622,clip]{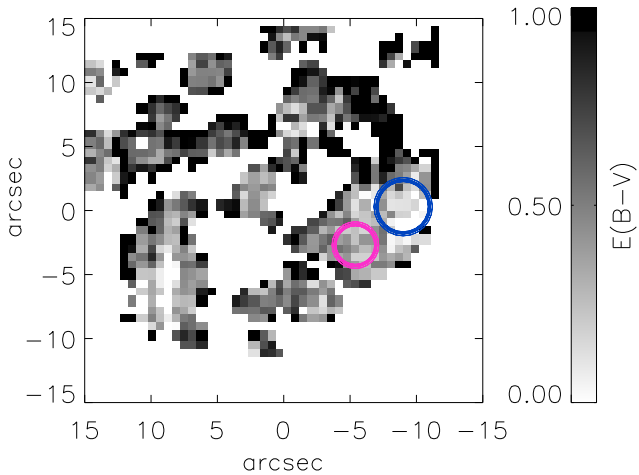}}
\caption{Reddening map, showing the value of \ebv. Half of the WR
  region at coordinates (--8.7, 0) appears relatively free of dust
  (white colour), while the SN region has \ebv = 0.63 mag. The small
  circle indicates the location of the SN region, and the larger
  circle the location of the WR region. Apart from the white
  colour in the WR region, white regions in this and following maps
  represent spaxels where the \halpha\ flux is smaller than
  3$\times10^{-17}$ erg~s$^{-1}$~cm$^{-2}$.}
\label{fig:ebv}
\end{figure}

The ratio of the maps \halpha/\hbeta\ provides a measure of the
reddening by dust. In the Case B recombination scenario, the
theoretical dust free ratio is 2.85 for a temperature of 10\,000~K
\citep{osterbrock89}. The reddening is calculated by the equation \(
\ebv =1.98
     [\log_{10}(\halpha_{\mathrm{obs}}/\hbeta_{\mathrm{obs}})-\log_{10}(2.85)]\).
     The factor 1.98 assumes a Milky Way type extinction law
     \citep{fitz99}, but choosing a different extinction curve does
     not change this factor significantly.  The reddening map is
     presented in Fig.~\ref{fig:ebv}. We adopted a threshold for the
     \halpha\ flux of 3$\times$10$^{-17}$~erg~s$^{-1}$~cm$^{-2}$ to
     calculate the extinction, because in regions where the flux is
     lower, the value of \ebv\ becomes correspondingly
     uncertain. Whereas the host has an overall \ebv = 0.30 mag,
     approximately half of the WR region appears to be dust free with
     an overall average of 0.17 mag.  On the other hand, the SN region
     appears to have significantly more reddening with an estimated
     average of \ebv = 0.63 mag.  A similar analysis can be completed
     using the \hbeta/\hdelta\ emission line ratio but this ratio has
     a larger uncertainty due to the fainter emission lines
     involved. The results obtained from this ratio agree with those
     for \halpha/\hbeta\ to within the uncertainties.  Using the tight
     correlation between the \halpha\ and 24$\mu$m luminosities
     \citep{kennicutt07} and the IR luminosity of 2.34$\times10^{40}$
     erg~s$^{-1}$ for the WR region \citep{lefloch06}, we measure an
     overall 3\% extinction correction of the \halpha\ flux. The
     spatial resolution of Spitzer is insufficient to resolve
     separately the SN region.

The reddening inferred from the FORS archive spectrum is \ebv = 0.16
and 0.14 mag for the WR and SN region, indicating that a significant
discrepancy exists only for the SN region. For this region,
\citet{sollerman05} measured \ebv=0 mag, while \citet{hammer06} found
\ebv = 0.39 mag and \citet{savaglio08} found $A_V=1.73$. If the
extinction in the SN region is as high as we find, this would imply
that the supernova was exceptionally (and unreasonably) bright and
blue.  If the extinction is larger than inferred from observations of
SN~1998bw, this could imply that the SN was in the foreground of a
dustier \ion{H}{ii} region.

The magnitudes of the SN region in the FORS images and the VIMOS data
agree to within 0.05 mag when the same aperture size is used.  Because
the long slit data cannot be used to calculate the magnitude as a
consistency check, we assume that the large extinction found in the
VIMOS data is correct, even though there is a significant disagreement
with some literature values.

\subsection{Star formation rates}
\label{sect:sfr}
\begin{figure*}
\centering
\begin{minipage}[c]{0.47\textwidth}
\resizebox{\hsize}{!}{\includegraphics[bb=48 360 412 622,clip]{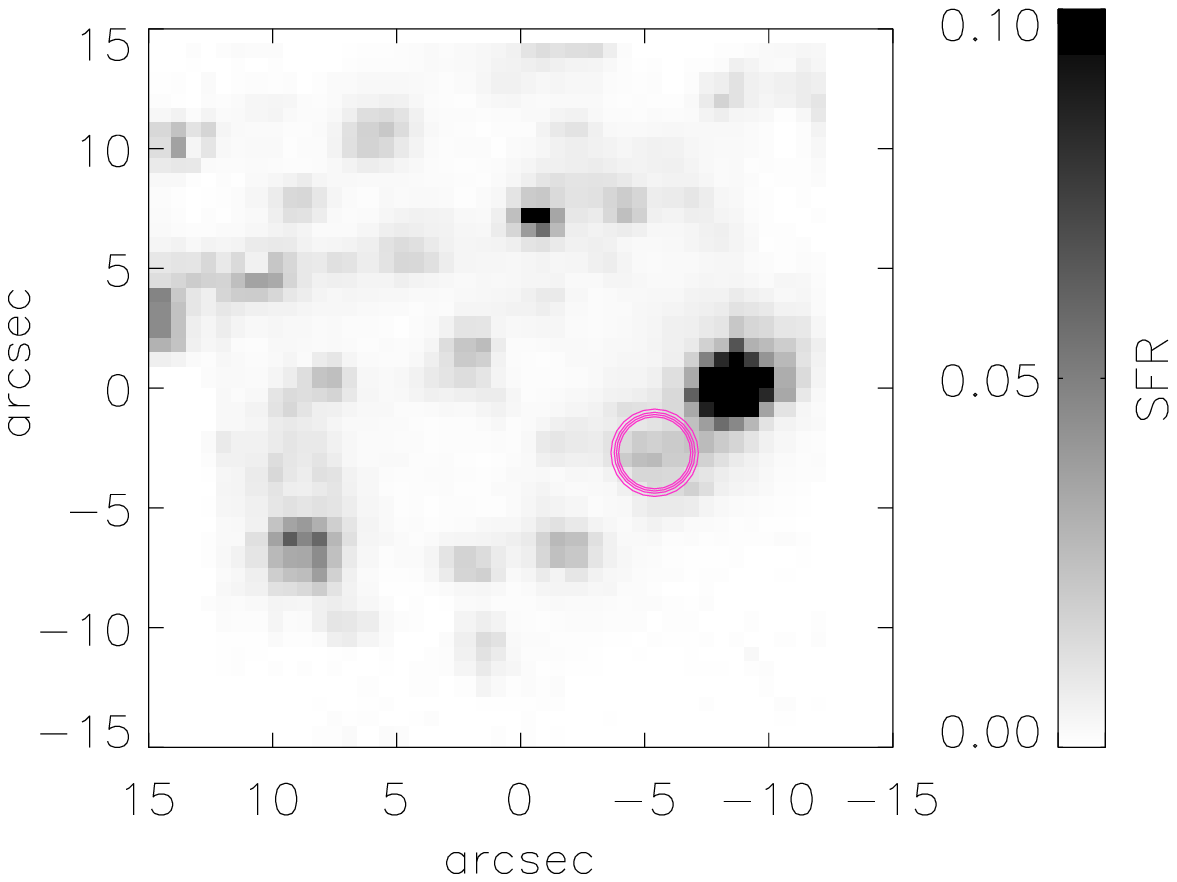}}
\end{minipage}%
\begin{minipage}[c]{0.47\textwidth}
\resizebox{\hsize}{!}{\includegraphics[bb=48 360 412 622,clip]{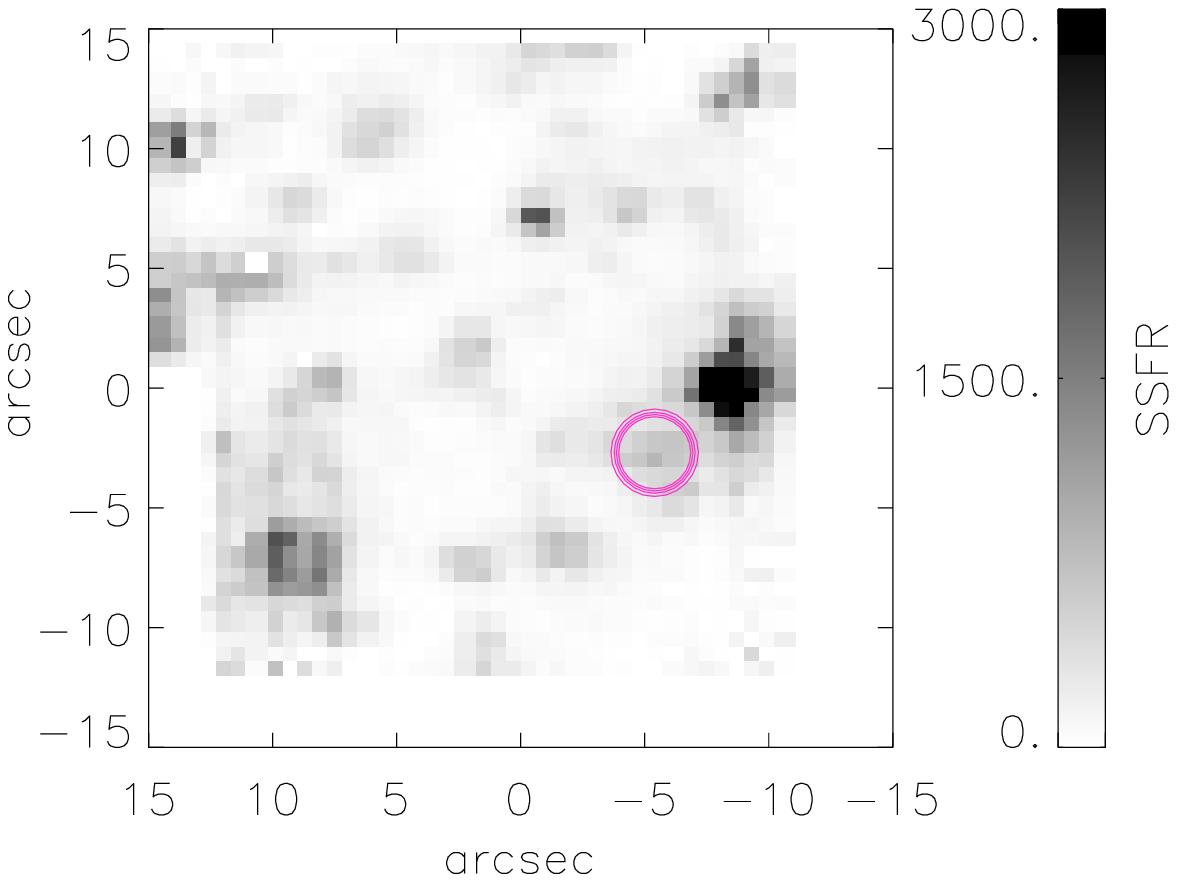}}
\end{minipage}
\caption{Left panel: Map of the SFR converted from the
  \halpha\ map. The units are in M$_{\odot}$~yr$^{-1}$~kpc$^{-2}$.
  Right panel: Map of the luminosity specific SFR in units of
  M$_{\odot}$~yr$^{-1}(L/L^*)^{-1}$~kpc$^{-2}$ (uncorrected for
  intrinsic extinction in the host). The WR region has the largest
  value over the face of the host. The circle marks the location of
  the SN region.}
\label{fig:sfr}
\end{figure*}

The \halpha\ emission line flux can be converted to a SFR given the
transformation in \citet{kennicutt98}.  This law was derived for
complete galaxy disks, and may be inappropriate for individually
resolved \ion{H}{ii} regions and very young regions experiencing
almost instantaneous bursts of star formation.  The ages derived for
individual \ion{H}{ii} regions in the GRB 1998bw host are similar, and
the \halpha\ luminosities are sufficiently high; we can therefore
assume that the conversion of \halpha\ luminosity to SFR is applicable
to individual regions \citep[see also][]{kennicutt07}.  A map of the
SFR surface density is shown in Fig.~\ref{fig:sfr}.

As derived from the \halpha\ emission line luminosity, the total SFR
within the FOV is 0.21~M$_{\odot}$~yr$^{-1}$. From the integrated
one-dimensional spectrum over the entire FOV, we measure
SFR=0.23~M$_{\odot}$~yr$^{-1}$. This difference is caused by
differences in the uncertainties in estimating the underlying
continuum: Within the image, the continuum is estimated in each
spaxel, while the continuum is measured in only one dimension in the
extracted spectrum.  Within the 30 arcsec square aperture, the
\halpha\ narrow-band image obtained by \citet{sollerman05} provides a
measurement of 0.22~M$_{\odot}$~yr$^{-1}$ (corrected for 10\% flux
from [\ion{N}{ii}] emission present in the narrow band too).

At $z$~$\sim$~1, GRB host galaxies have typical UV-based specific star
formations rates (SSFRs) of about
10~M$_{\odot}$~yr$^{-1}(L/L^*)^{-1}$, where the luminosities $L$ are
normalised to the $B$ band luminosity $L^*$ corresponding to the
magnitude $M^*~=~-21$ \citep{christensen04b}.  To determine the SSFR
of the GRB~980425 host, the spectra in the blue high resolution cube
are convolved with a Bessell transmission function \citep{bessell90}
to derive the $B$ band magnitude in each spaxel. The SFR map is then
divided by the relative luminosity $L/L^*$ as shown in the right panel
of Fig.~\ref{fig:sfr}, which measures the \halpha\ flux relative to
the blue continuum luminosity. In the WR region, the \halpha\ flux is
much larger for a given continuum flux than in other regions.

Since the SFR measurement is more relevant on a galaxy wide scale, we
also analyse the integrated properties of the galaxy. The integrated
magnitude in the VIMOS data is $m_B$=15.3 mag, corresponding to an
absolute magnitude of $M_B=$--17.5, and
SSFR=5.3~M$_{\odot}$~yr$^{-1}(L/L^*)^{-1}$.  This is slightly smaller
than the UV-based SSFRs of GRB hosts at higher redshift. The entire
GRB host has a SSFR$\approx$7~M$_{\odot}$~yr$^{-1}(L/L^*)^{-1}$
\citep{sollerman05}. If we apply a correction for the internal
extinction in the host (\ebv=0.3) we derive
SSFR=10.2~M$_{\odot}$~yr$^{-1}(L/L^*)^{-1}$.

The high value for the WR region is distinctive compared to that of
GRB hosts at $z\sim1$. However, the explanation is that the individual
star forming regions are resolved (here normalised to kpc$^{-2}$),
while the higher redshift hosts are investigated using integrated
properties. Other lower redshift GRB host with high SSFRs include the
GRB 030329 host with $\sim$25~M$_{\odot}$~yr$^{-1}(L/L^*)^{-1}$
\citep{gorosabel05}, and the GRB 031203 host with
SSFR$\sim$39~M$_{\odot}$~yr$^{-1}(L/L^*)^{-1}$ \citep{sollerman05}.
All of these values are derived from emission line based SFRs, and it
is known that there are discrepancies between the SFRs determined from
the UV flux and from emission lines, albeit the discrepancies found
are typically lower by a factor of less than 2. Nevertheless, the high
values for GRB hosts are unusual compared to field galaxies, which
have typical UV continuum based
SSFR~$\lesssim$~5~M$_{\odot}$~yr$^{-1}(L/L^*)^{-1}$
\citep{christensen04b}.

\subsection{Electron temperature and densities}
\label{sect:dens}
Ideally, the temperatures and densities of the \ion{H}{ii} regions
should be derived by including faint nebular emission lines such as
the [\ion{O}{iii}]~$\lambda$4363, or the [\ion{N}{ii}]~$\lambda$5755
line \citep{osterbrock89}.  Only for one region do we detect the
temperature sensitive [\ion{O}{iii}]~$\lambda$4363 line (right panel
in Fig.~\ref{fig:maps}).  In the FORS data, we also detect the faint
line only in the WR region.  From the ratio O3=(4959+5007)/4363 we
find that T=10\,300$\pm$1000~K for the WR region. In contrast,
\citet{hammer06} measured temperatures for the SN and WR region of
between 11\,000 and 12\,000~K.

To determine the electron densities, we analyse the ratio of the
[\ion{S}{ii}]~$\lambda\lambda$6717,6731 lines using the IRAF package
NEBULAR, which is based on the code presented by \citet{shaw95}. At
the spectral resolution of VIMOS ($R\sim$2900 as measured from sky
lines), the sulfur lines are clearly resolved, while at the lower
resolution of the FORS spectrum ($R$=650) the lines are blended. The
WR region has a density of 120 cm$^{-3}$ derived from the
[\ion{S}{ii}] line ratio, which for most regions is close to 1.3.
Since the flux calibration does not affect the uncertainty in the
ratio, the error for the ratio is 0.03 based on the number counts in
the spectra. A density of the order 100~cm$^{-3}$ is typically derived
for other GRB hosts \citep{prochaska04,wiersema07,savaglio08}.

\subsection{Abundances}
\label{sect:metal}
\begin{figure}
\resizebox{\hsize}{!}{\includegraphics[bb=48 360 412 622,clip]{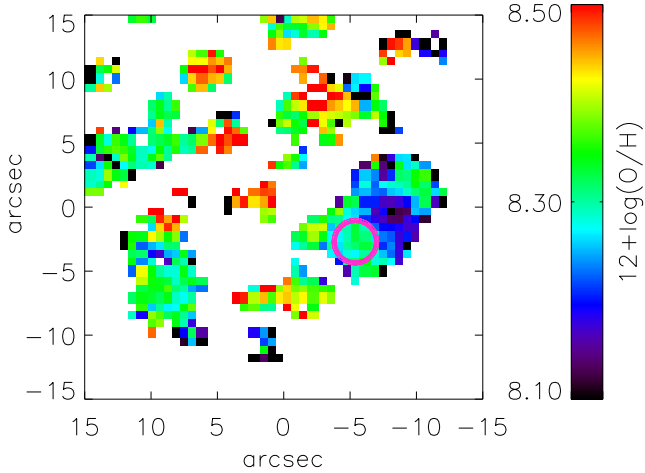}}
\caption{Oxygen abundance map derived from the O3N2 abundance
  diagnostics. The WR region has the lowest abundance in the whole
  galaxy, and the SN region has the second lowest abundance although
  similar to other regions within the 1$\sigma$ uncertainty of 0.14
  dex. The circle indicates the location of the SN region. [{\it See
      the electronic edition of the Journal for a colour version of
      this figure}.]}
\label{fig:o3n2}
\end{figure}

We use the O3N2 abundance diagnostics from \citet{pettini04} to
determine the oxygen abundance. This diagnostics involve the line flux
ratios [\ion{O}{iii}]~$\lambda$5007/\hbeta\ and
[\ion{N}{ii}]~$\lambda$6586/\halpha, whose lines are close in
wavelength so that extinction plays a minor role.
Figure~\ref{fig:o3n2} shows the metallicity map of the host galaxy and
illustrates the values in Table~\ref{tab:emlines} in the units
12+$\log$(O/H), with the solar value of 8.66 adopted from
\citet{asplund04}. Extracting the one-dimensional spectra of the WR
and SN regions yields 12+log(O/H)=8.16 and 8.30, respectively, which
corresponds to 0.32 and 0.44 solar abundance.  The uncertainties in
the abundance is 0.03 dex if we propagate the uncertainties in the
emission line fluxes. In comparison, the 1$\sigma$ scatter uncertainty
of 0.14 dex for data points in \citet{pettini04} used to derive the
O3N2 diagnostics is clearly far higher. The calibration is derived for
values of O3N2$<$1.9, which is valid for all regions in the GRB host.
Considering the large scatter of the calibration, the derived
metallicities in Table~\ref{tab:emlines} are similar to within a
margin of error of 3$\sigma$ and the WR and SN region metallicities
are consistent to within a margin of error of 1$\sigma$.

While the value for the WR region is lower than for other regions, it
is also striking that the SN region has the second lowest metallicity;
within the margins of uncertainty, many other regions, however, have
the same abundance. Since these metal poor regions have brighter
emission lines than the remaining regions, they dominate the
integrated spectrum of the host, even though some values in
Table~\ref{tab:emlines} correspond to higher abundances (0.6--0.8
solar).  The total galaxy spectrum has 12+log(O/H)=8.27 (0.40 solar),
i.e.  similar to that of the SN region.

Using the N2 abundance diagnostics from \citet{pettini04} the same
values of abundances are derived to within the uncertainties, while
the S2 diagnostics, which use the
\halpha/[\ion{S}{ii}]~$\lambda\lambda$6717,6731 ratio and the O3N2
calibration in \citet{yin07} provides systematically higher abundances
by 0.1--0.15 dex for all regions. The calibration in \citet{yin07} was
derived for a larger number of galaxies than in \citet{pettini04},
where the calibration was derived from \ion{H}{ii} regions, for which
the abundances are measured using temperature sensitive lines. It is
well known that different diagnostics can measure different values of
abundances \citep{kewley08}, and here we do not attempt to correct for
this effect. The important point for the GRB host is not the absolute
values in metallicity, but the variations of the metallicity over the
galaxy image, and that one particularly bright region can dominate the
measurement of the integrated abundance.

In the WR region, where the temperature sensitive
[\ion{O}{iii}]~$\lambda$4363 line is detected with a flux of
$\sim$~5~$\times$~10$^{-16}$~erg~s$^{-1}$~cm$^{-2}$, the equations in
\citet{izotov06} and the de-reddened fluxes imply an oxygen abundance
of 12+log(O/H)=8.53$\pm$0.10, which is significantly larger than that
derived for the O3N2 diagnostics. This discrepancy suggests that the
O3N2 diagnostics may not be a valid approximation for this particular
galaxy, and that the abundance may even be on average close to solar
in contradiction to the general idea that GRB progenitors and their
hosts have sub solar metallicities.

The strong emission lines in Table~\ref{tab:emlines} can be used to
derive the (N/O) abundance ratio.  To calculate this ratio, we added
the contributions to the abundance O/H$^+$=O$^+$/H$^+$+O$^{
  2+}$/H$^+$, and N was obtained by multiplying the N$^+$/H$^+$ ratio
with an ionisation correction factor using the equations in
\citet{izotov06}.  The results are presented in the final column of
Table~\ref{tab:emlines}.  For the different \ion{H}{ii} regions, a
constant temperature of 10$^4$ K is assumed.  This is the strongest
assumption that can produce a ratio 0.2 dex lower, if the temperature
is 2000~K lower. Correspondingly, a 0.2 dex higher ratio would require
the temperatures to be higher by $\sim$3000 K. However, such a high
electron temperature is unexpected since most regions are not
particularly oxygen poor. A low temperature ($<$8000~K) would imply a
N/O ratio lower than found for most metal poor galaxies (log(N/O)=
--1.8).  The uncertainties in the line fluxes imply an uncertainty of
0.1 dex in log(N/O).  Since the densities of the various regions are
close to 100~cm$^{-1}$, the small differences do not imply a
significant correction to the N/O ratio. Overall this implies an
uncertainty in the N/O ratio of 0.2 dex.

The N/O ratio can be used to probe the star formation history of the
galaxy given the different timescales for the primary nitrogen
production and the secondary contribution of nitrogen from
intermediate mass stars. Compared to the solar value of
log(N/O)$_{\odot}$=--0.81 \citep{holweger01}, most of the regions have
sub solar values.  Within the uncertainties the WR and SN regions have
similar values of N/O, while the value for the entire galaxy is again
dominated by these bright \ion{H}{ii} regions.  Taken at face value,
the N/O ratios indicate that the WR region is the youngest, but not as
young as the suggested by the N/O ratio for some damped Lyman-$\alpha$
systems \citep{centurion03,pettini08}. Contamination from previous
generations of stars to gas over timescales longer than 100--300 Myr
in the WR region is responsible for this effect.  The presence of WR
stars is expected to increase the N/O ratio \citep{meynet05}, although
\citet{izotov06} found that the enhancement is small.  For GRB hosts,
it remains unclear whether an enhancement of nitrogen is present
because the measurement errors are large \citep[see][]{wiersema07}.

\subsection{Stellar population modelling}
\label{sect:popul}

The observed spectra were fit by the spectral templates derived from
simple stellar evolution models from Starburst99 \citep{leitherer99},
which included both stellar and nebular emission.  We chose to fit the
VIMOS spectra with models of 0.4 solar metallicity because this is the
most representative value for the host galaxy. The models assume an
instantaneous burst of star formation, and a Salpeter law initial mass
function between 1 and 100 M$_{\odot}$.  The best-fit templates are
determined from a standard $\chi^2$ fitting technique.  Using the
intrinsic extinction in Table~\ref{tab:emlines} to de-redden the
spectra before fitting the models, we find that the best-fit age of
6~Myr for the WR region, whereas the best-fit age is 3~Myr for the SN
region due to the large intrinsic extinction.  For all \ion{H}{ii}
regions, the best-fit ages are between 2 and 10~Myr as listed in Col.
4 in Table~\ref{tab:deriv}. With the smaller intrinsic reddening
inferred from the SN region, \citet{sollerman05} found an age of 6 Myr,
and a corresponding progenitor mass of 30$\pm$5~M$_{\odot}$.

For comparison, we also fitted the de-reddened spectra from individual
regions to the high resolution spectral templates from
\citet{bruzual03}. The models assumed a single burst population and a
Salpeter law initial mass function and again a 0.4 solar
metallicity. We generally found a larger scatter between individual
\ion{H}{ii} region with best fit ages between 1 and 200 Myr. Since
these models did not include the nebular emission component, which is
dominant for the very young regions, the ages determined from the
Starburst99 model should represent the regions more appropriately.
 
The ages derived from the individual regions are in broad agreement
with those derived from the equivalent widths in Sect~\ref{sect:ew},
which indicate ages of 3 to 6 Myr for the individual \ion{H}{ii}
regions, and the emission line ratios in Sect.~\ref{sect:ratios}.
Discrepancies in the individual age estimates are probably caused by
uncertainties in the reddening estimate, which affect the spectral
fittings, while the EWs are largely unaffected by the reddening.  The
continuum provides an insight into the older populations of stars,
while the emission lines represent the most recent star forming
episode. Since the younger regions have higher extinctions, this could
affect the EWs. Irrespectively of this fact, both the EWs and the
population fittings indicate young ages for all regions ($<$10 Myr).

\subsection{Mass estimates}
\label{sect:mass}
The best-fit Starburst99 models were used to estimate the masses of
the different \ion{H}{ii} regions. Since the template models were
created for stellar populations of $10^6$~M$_{\odot}$, a simple
scaling to the observed spectrum was required to derive the mass.
Using the first entries in Table~\ref{tab:emlines}, the WR and SN
regions have similar masses of 0.44$\times10^6$~M$_{\odot}$ and
0.5$\times10^6$~M$_{\odot}$, respectively, while the mass of the
luminous stars in the galaxy is $2\times10^7$~M$_{\odot}$.
Uncertainties in the mass estimate includes both the error for the
absolute flux, and the correction for the reddening yielding a total
uncertainty of 20\%.

In Table~\ref{tab:deriv}, the masses are used with estimates of the
SFR to derive additional parameters, such as mass-to-light ratios and
specific SFRs, using mass as a means of normalization instead of the
luminosity for the various regions. We note that the units in Cols.  7
and 8 are different.  We checked the mass estimates for the individual
regions by comparing the FORS $R$ band magnitudes, extracted within
the same aperture as the VIMOS spectra, with the model predictions of
the 0.4 solar metallicity \citet{bruzual03} models.  We found a
generally good agreement with the values listed in
Table~\ref{tab:deriv}.  The derived mass depended strongly on the
assumed age for the regions and the extinction applied.

Comparing the mass SSFR with the definition of the luminosity SSFR
used in Sect.~\ref{sect:sfr} (the two final columns in
Table~\ref{tab:deriv}), the values agree in general to within a factor
of a few. The determination of the `mass SSFR' involves additional
uncertainties because of the model dependencies of the mass-to-light
ratio and age determination. These uncertainties do not influence the
luminosity SSFR since this quantity only includes directly observed
parameters and is therefore an excellent proxy for the more generally
used mass SSFR. The SSFRs in the Table are higher compared to those
derived for higher redshift GRB hosts, which have typical
extinction-corrected luminosity SSFRs = 1--30~M$_{\odot}$~yr$^{-1}$
$(L/L^*)^{-1}$ \citep{christensen04b}. The reason is that the
extinction determined from the emission line ratios in the \ion{H}{ii}
regions was far higher than found from the integrated spectral energy
distributions, and the correction applied to the SFRs was therefore
correspondingly higher.

\begin{table*}
\centering
\begin{tabular}{l r@{.}l r@{.}l r r@{.}l r@{.}l r@{.}l r@{.}l}
\hline
\hline
\noalign{\smallskip}
reg. &  \multicolumn{2}{c}{SFR} &  \multicolumn{2}{c}{SFR (corr.)} & Age &  \multicolumn{2}{c}{Mass} &  \multicolumn{2}{c}{M/$L_B$} &  \multicolumn{2}{c}{Mass SSFR (corr.)} &  \multicolumn{2}{c}{Lum SSFR (corr.)}\\
     & \multicolumn{2}{c}{($\times10^{-3}$~M$_{\odot}$~yr$^{-1}$)} 
 & \multicolumn{2}{c}{($\times10^{-3}$~M$_{\odot}$~yr$^{-1}$)} & (Myr) 
& \multicolumn{2}{c}{($\times10^6$~M$_{\odot}$)} &
\multicolumn{2}{c}{(M$_{\odot}/L_{B,{\odot}}$)} &
\multicolumn{2}{c}{(M$_{\odot}$~Gyr$^{-1}$ M$_{\odot}^{-1}$)} & \multicolumn{2}{c}{(M$_{\odot}$~yr$^{-1}$ $(L/L*)^{-1}$)} \\
(1) &  \multicolumn{2}{c}{(2)} & \multicolumn{2}{c}{(3)} & (4) & \multicolumn{2}{c}{(5)} & \multicolumn{2}{c}{(6)} & \multicolumn{2}{c}{(7)} & \multicolumn{2}{c}{(8)}\\ 
\noalign{\smallskip}
\hline 
\noalign{\smallskip}
Galaxy & 233&0 & 447&0 & 2 & 20&1   & 0&014 & 22&2 & 10&9 \\
WR reg.&  78&3 & 113&3 & 2 &  0&44  & 0&007 & 257& & 58&2\\
SN reg.&  5&84 & 22&9  & 3 &  0&50  & 0&035 & 45&8 & 54&3\\
\noalign{\smallskip}
\hline
\noalign{\smallskip}
WR  & 54&5&  58&0   & 6 & 0&28& 0&007 & 207&  & 48&9\\
SN  & 2&52&  12&9   & 3 & 0&20& 0&035 & 64&3  & 76&4\\
    & 1&22&  20&1   & 3 & 1&15& 0&149 & 17&5  & 89&1\\
    & 1&20&  5&25   & 3 & 0&18& 0&057 & 29&2  & 56&8\\
    & 1&01&  5&48   & 3 & 0&43& 0&060 & 12&7  & 26&1\\
    & 2&09&  5&32   & 4 & 0&22& 0&020 & 24&2  & 16&7\\
    & 1&45&  4&99   & 4 & 0&19& 0&033 & 26&2  & 29&6\\
    & 2&30&  26&1   & 3 & 0&46& 0&096 & 56&8  & 185&1\\
    & 1&02&  3&45   & 3 & 0&28& 0&028 & 12&3  & 11&7\\
    & 6&31&  12&6   & 6 & 0&20& 0&012 & 63&2  & 26&3\\
    & 0&98&  3&61   & 3 & 0&35& 0&034 & 10&3  & 12&0\\
    & 0&68&  2&92   & 3 & 0&26& 0&041 & 11&2  & 15&7\\
    & 2&24&  4&49   & 3 & 0&13& 0&010 & 34&5  & 12&3\\
    & 1&43&  20&6   & 3 & 0&46& 0&122 & 44&8 & 185&3\\
    & 1&71&  3&73   & 10 & 0&57& 0&038& 6&55  & 8&43\\
    & 2&09&  3&08   & 10 & 0&29& 0&020& 10&6  & 7&09\\
    & 2&07&  2&07   & 3 & 0&08& 0&005 & 25&9  & 4&39\\
    & 5&58&  5&58   & 2 & 0&07& 0&012 & 79&8  & 32&0\\
    & 1&41&  2&64   & 6 & 0&10& 0&007 & 26&4  & 6&13\\
    & 3&52&  3&52   & 3 & 0&12& 0&005 & 29&4  & 4&58\\
    & 0&48&  1&27   & 3 & 0&11& 0&017 & 11&6  & 7&48\\
    & 1&20&  1&20   & 3 & 0&04& 0&004 & 30&1  & 4&19\\
    & 0&95&  0&95   & 3 & 0&03& 0&004 & 31&5  & 4&80\\
\noalign{\smallskip}
\hline
\noalign{\smallskip}
mean & 4&$\pm$11  &  9&$\pm$13 & 4$\pm$2  &0&3$\pm$0.2 &
0&04$\pm$0.04 & 38&$\pm$42 & 38&$\pm$52\\
\noalign{\smallskip}
\hline
\end{tabular}
\caption{Table for derived quantities for the different regions. The
  last row gives the mean and standard deviation for all the
  regions. Column 2 is the SFR derived from the observed
  \halpha\ flux, and col. 3 corrects the SFR for the extinction given
  in the Table~\ref{tab:emlines}. Column 4 gives the best fit age. The
  mass (col. 5) is determined from population modelling. The mass to
  light ratio (Col. 6) uses the luminosity $L_B$ derived from the B
  magnitude relative to the AB magnitude for the sun
  M$_{B,{\odot}}$=5.33. Column 7 lists the specific SFR per Gyr
  normalised to the mass after correction for the host extinction, and
  Col. 8 lists the luminosity specific SFR, both derived after
  correcting for intrinsic extinction.}
\label{tab:deriv}
\end{table*}

For the entire galaxy, we compared the mass derived from the stellar
population model fitting to masses derived from the absolute
magnitudes measured from the FORS $R$ band image.  Within the VIMOS
FOV, the ratio of the luminosity is $L/L_{\odot}=10^{9.3}$. Assuming a
mass-to-light ratio of 0.1 \citep[calibrated using the $K$ band as
  appropriate for GRB hosts cf.][]{savaglio08}, the mass within the
field was 2$\times10^8$ M$_{\odot}$. In comparison, the mass estimated
from the kinematical data was a factor of three times higher. We
tested this mass estimate with the spectral energy distribution
fitting codes Z-Peg \citep{leborgne02}, and HyperZ \citep{bolzo00} on
the measured broad band magnitudes. Both SED fitting codes provide
similar measurements of mass to within the uncertainties.

To investigate whether the galaxy is unusual in comparison with field
galaxies, we compared our SFR measurement with SFRs measured for SDSS
galaxies.  The total stellar mass of the host galaxy that we study,
infers an expected star formation rate of 0.3~M$_{\odot}$~yr$^{-1}$
following the distribution values for galaxies in the SDSS
\citep{brinchmann04}. The \halpha\ flux corrected for the host
extinction in Sect.~\ref{sect:sfr} is 0.4~M$_{\odot}$~yr$^{-1}$, which
is within the distribution of SFRs for SDSS galaxies of similar
mass. With this SFR and a total mass of 2$\times$10$^8$~M$_{\odot}$,
the specific SFR for the host
(SSFR=2.0~M$_{\odot}$~Gyr$^{-1}$~M${_\odot}^{-1}$) is located at the
upper end of the distribution for SDSS galaxies \citep[see Fig. 24
  of][]{brinchmann04}.

\subsection{Variation of properties in the \ion{H}{ii} regions}
To illustrate the various average properties in
Table~\ref{tab:emlines} and ~\ref{tab:deriv} relative to the values
measured for the SN and WR regions, we calculate for each quantity the
difference from the average (written here for the SFR):

\[\mathrm{Deviation}= \frac{ \mathrm{SFR(SN)} - \langle
\mathrm{SFR(\ion{H}{ii})} \rangle  }{ \sigma
  (\mathrm{SFR(\ion{H}{ii})})},\]
where $\langle \mathrm{SFR(\ion{H}{ii})} \rangle$ is the mean value
for the SFR, and $\sigma (\mathrm{SFR(\ion{H}{ii})})$ is the standard
deviation as listed in the last row in Table~\ref{tab:emlines} and
~\ref{tab:deriv}.

\begin{figure}
\resizebox{\hsize}{!}{\includegraphics[bb=87 343 536 700,clip]{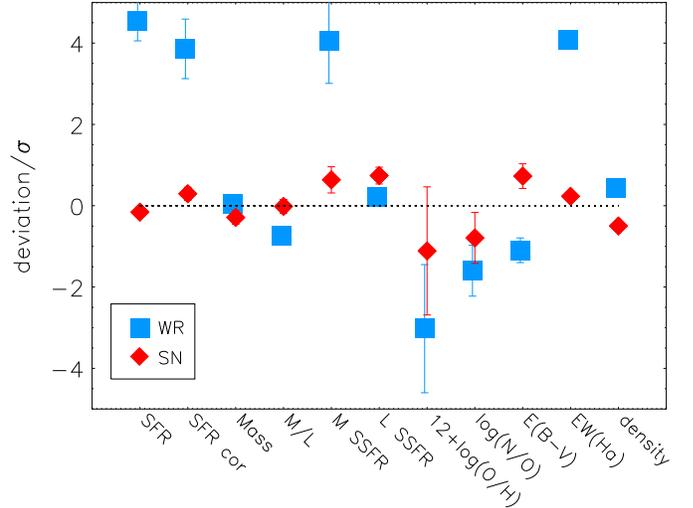}}
\caption{Deviation from the average of the properties listed on the
  x-axis. The WR region is indicated by squares and the SN region by
  diamonds.  Error bars are in some cases smaller than the symbols.}
\label{fig:dev}
\end{figure}

Figure~\ref{fig:dev} shows this quantity for all other
properties. Notably the SFRs (corrected and uncorrected for the host
extinction) of the WR region are significantly (4 $\sigma$) above the
average for all \ion{H}{ii} regions, and likewise the abundances are
lower for both the WR and SN regions, although the values are
consistent with the average within the uncertainties.  Since the
masses of the WR and SN regions are similar, it follows that the
extinction-corrected mass SSFR is also much higher for the WR
region. On the other hand, the luminosity SSFRs for both the SN and WR
regions appear to be only slightly above the average.  Since the $B$
band magnitude of the SN region is much fainter than for the WR
region, and no correction for extinction is applied to the magnitude
before the luminosity SSFRs are calculated, the luminosity weighted
SSFR for the two regions are similar and close to the average for the
\ion{H}{ii} regions. For none of the parameters does the figure change
significantly if the median value is used in the calculation instead
of the average.

The difference compared to the values in the SSFR map in
Fig.~\ref{fig:sfr} is due to the fact that the map is uncorrected
for internal extinction in the host, whereas the SSFRs in the Tables
and in Fig.~\ref{fig:dev} are extinction corrected.

\section{Emission line ratios, ages, and metallicities}
\label{sect:ratios}
While the SN region in the GRB 980425 host is the second most
metal-poor region, it is not dust-free and most properties (SFRs, EWs,
masses) are similar to those of the remaining \ion{H}{ii} regions in
the host galaxy.  In contrast, the WR region appears to be
characteristic of that expected for a GRB region in terms of these
properties \citep[see][]{christensen04b}.

\begin{figure*}[t!]
\begin{minipage}[c]{0.5\textwidth}
\resizebox{\hsize}{!}{\includegraphics[]{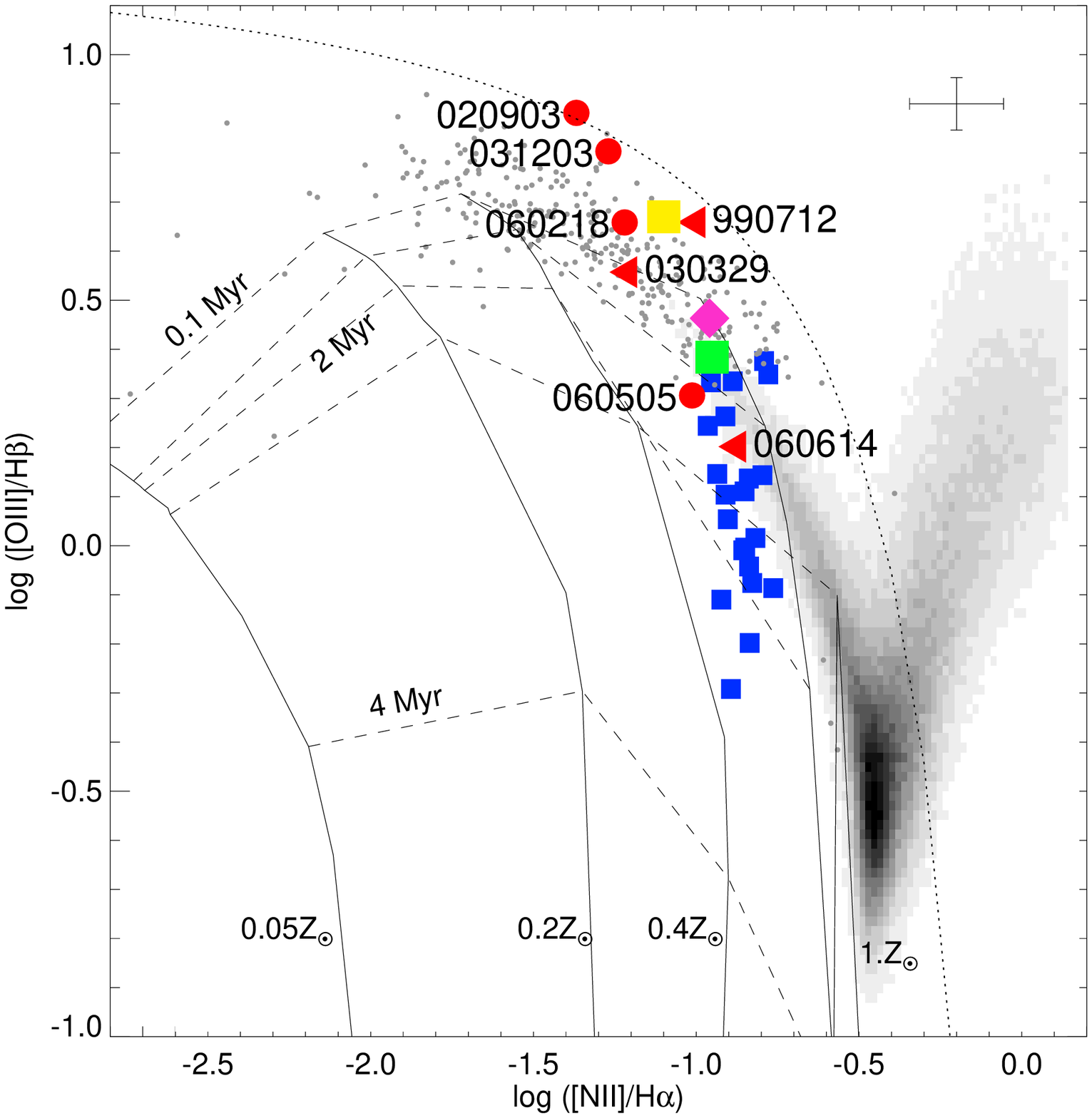}}
\end{minipage}%
\begin{minipage}[c]{0.5\textwidth}
\resizebox{\hsize}{!}{\includegraphics[]{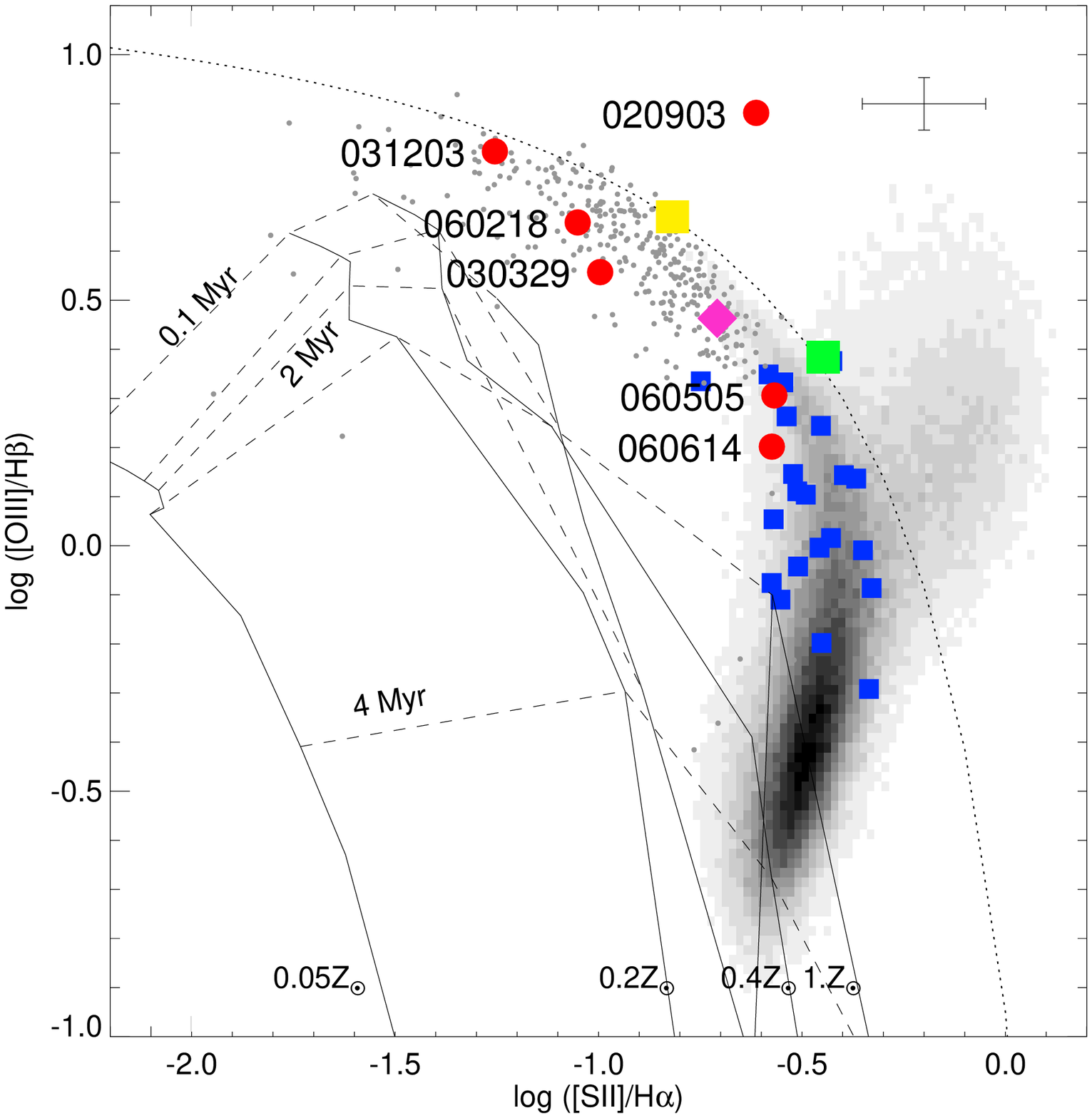}}
\end{minipage}
\caption{Emission line ratios for GRB hosts:
  [\ion{O}{iii}]/\hbeta\ vs.  [\ion{N}{ii}]/\halpha\ (on the left) and
  vs. [\ion{S}{ii}]/\halpha\ (on the right).  Individually resolved
  \ion{H}{ii} regions are shown as small blue squares, the SN region
  (larger green square), and the WR region (large yellow square), and
  the entire galaxy (pink diamond).  The median error bar for the
    sample is indicated in the upper right corner.  The distinct
  location of the WR and SN regions indicate that they are special
  relative to other \ion{H}{ii} regions.  The evolutionary models that
  link emission line regions at ages from 0.1 to 5 Myr for 0.05, 0.2,
  0.4 and solar metallicities are taken from \citet{dopita06}. The
  dotted, curved lines denote the separation of pure star forming
  galaxies to AGN dominated emission line ratios
  \citep{kewley01,kauffmann03}.  To compare with field galaxies, the
  gray scale area represent SDSS galaxies \citep{tremonti04}, and the
  small grey dots SDSS galaxies with metallicities between 0.03 and
  0.7 solar \citep{izotov06}.  Emission line ratios for other GRB
  hosts are overlayed by the named red circles. Two hosts have upper
  detection limits for [\ion{N}{ii}] and are represented by red
  triangles (right hand limits); GRB~030329 \citep{gorosabel05} and
  GRB~990712 \citep{christensen04a}.  Other references: GRB990712:
  \citet{yoldas06}, GRB~031203: \citet{prochaska04}, GRB~020903:
  \citet{hammer06}, GRB~060218: \citet{wiersema07}, GRB~060614:
  \citet{gal-yam06}, GRB~060505: \citet{thoene08}. A database is
  available at the GHostS web site ({\tt www.grbhosts.org}). {\it See
    the electronic edition of the Journal for a colour version of this
    figure}. }
\label{fig:n2}
\end{figure*}

\begin{figure*}[t!]
\begin{minipage}[c]{0.5\textwidth}
\resizebox{\hsize}{!}{\includegraphics[]{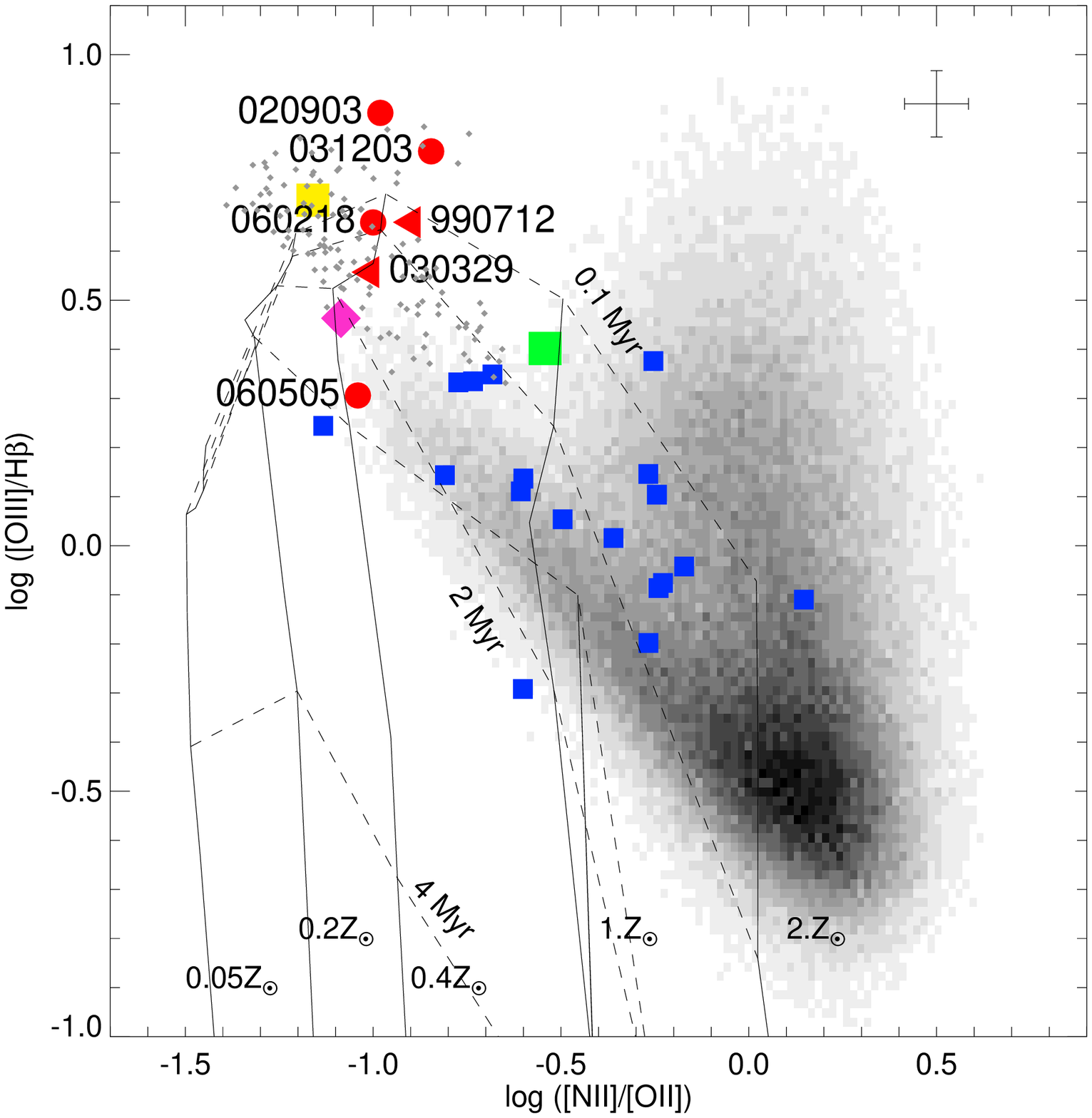}}
\end{minipage}%
\begin{minipage}[c]{0.5\textwidth}
\resizebox{\hsize}{!}{\includegraphics[]{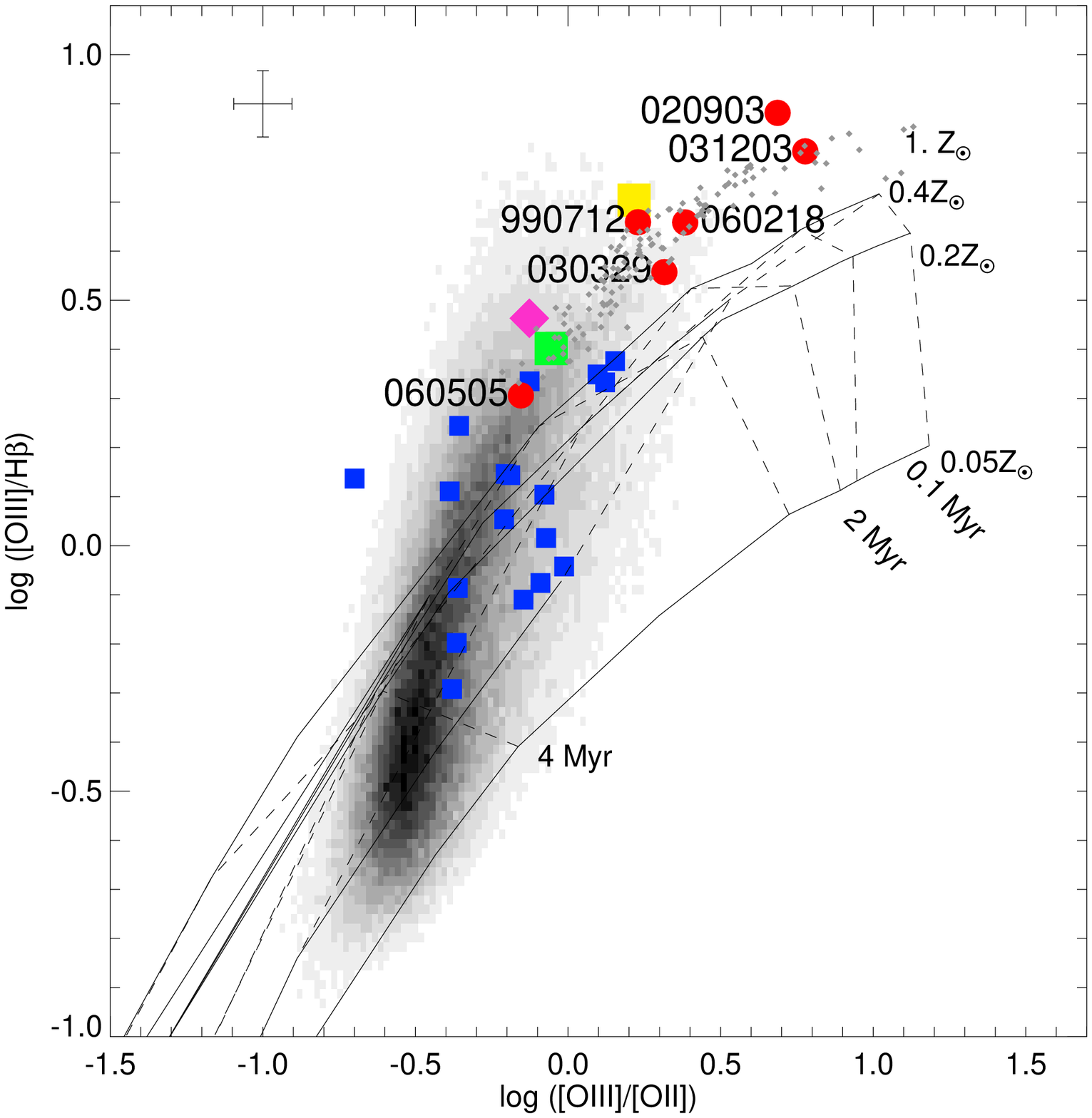}}
\end{minipage}
\caption{Emission line ratios [\ion{O}{iii}]/\hbeta\ vs.
  [\ion{N}{ii}]/[\ion{O}{ii}] (on the left) and
  vs. [\ion{O}{iii}]/[\ion{O}{ii}] (on the right).  The symbol shapes
  and colours are the same as in Fig.~\ref{fig:n2}.}
\label{fig:o2}
\end{figure*}

We compare the emission line ratios
[\ion{O}{iii}]/\hbeta\ vs. [\ion{N}{ii}]/\halpha\ and
[\ion{S}{ii}]/\halpha\ to other GRB hosts and other star-forming
galaxies in Fig.~\ref{fig:n2}. The individual regions (uncorrected for
the internal extinction in the host) from Table~\ref{tab:emlines} are
represented by the small squares.  De-reddening the fluxes should not
affect the line ratios significantly.  The emission line ratios are
compared with those of other GRB host found in the literature. The
emission line ratios were usually obtained from integrated spectra of
the hosts.

Figure~\ref{fig:n2} also includes the theoretical emission line ratios
for models of \ion{H}{ii} region of various metallicities and ages
\citep{dopita06}. Instead of the frequently used ionization parameter,
the models involve a parameter, $R$, which represents the fraction of
the star cluster mass and the pressure in the interstellar medium. For
typical stellar masses of the \ion{H}{ii} regions between
10$^5$--10$^6$~M$_{\odot}$ in Table~\ref{tab:deriv}, and a temperature
of 10$^4$~K and a density of 100~cm$^{-3}$ in Sect.~\ref{sect:dens},
this implies that $-1<\log R<0$. The tracks in Fig.~\ref{fig:n2}
assume the fraction $\log R=0$. For different choices of $R$, the model
ages of the \ion{H}{ii} regions differ slightly \citep[for details
  see][]{dopita06}, but this will not change the conclusion for the
relative ages of the GRB host emission line regions.

Except for GRB~060505 and GRB~060614, all GRB hosts are located at or
above the 0.1 Myr model, with their model metallicities between 0.4
and 1 solar. The offset between the models and the data points in the
right hand panel, which use the [\ion{S}{ii}]/\halpha\ ratio, is real,
and probably due to a deficiency of the models because they do not
include emission from a diffuse radiation field, which would increase
the [\ion{S}{ii}] flux.  Furthermore, the models of different
metallicities for [\ion{S}{ii}] provide prediction that are close to
each other, whereas they are distinctly separated in the [\ion{N}{ii}]
panel. Hence, the [\ion{N}{ii}]/\halpha\ line ratio is a more
effective tool for discriminating between metallicities than the
[\ion{S}{ii}]/\halpha\ line ratio. This may explain why the data for
the \ion{H}{ii} regions are clearly separate from the SDSS galaxies in
the left panel, because they are more metal poor than the SDSS
galaxies, while their data coincide in the right hand panel.  The
important point is that in this plot the GRB regions are offset
towards the higher [\ion{O}{iii}]/\hbeta\ ratio values (smaller age)
but still have a range of metallicities. Figure~\ref{fig:o2} presents
a similar plot for the line ratios:
[\ion{O}{iii}]/\hbeta\ vs. [\ion{N}{ii}]/[\ion{O}{ii}] and
[\ion{O}{iii}]/[\ion{O}{ii}]. We note that, due to the relatively
large uncertainties in the [\ion{O}{ii}] line fluxes, the scatter is
larger, but trends similar to those in Fig.~\ref{fig:n2} are evident.

It is interesting to note that in both panels of Fig.~\ref{fig:n2} and
Fig.~\ref{fig:o2}, the WR data occupy a similar region of the plot as
most of the other GRB hosts, which are significantly offset towards
the top. Similarly the SN region is located towards the top relative
to the other \ion{H}{ii} regions in the host. This would indicate that
GRBs occur in significantly younger regions -- at least in the low
redshift hosts where the emission line ratios can be studied.

The GRB~060505 host was also studied by resolved spectroscopy i.e. by
long slit observations \citep{thoene08}. The symbols in
Figs.~\ref{fig:n2}, and \ref{fig:o2} correspond to the emission line
ratios for the region where the GRB occurred. Emission lines from four
other regions are analysed, and although we do not show the individual
regions for this host here, the GRB region is also offset to the top
of the figure relative to the other regions. In the GRB~060505 host,
this implies that the GRB occurred in the youngest region probed by
the long slit spectrum.

We also compare the emission line ratios with the integrated ones for
field galaxies in the SDSS. The gray scale background images represent
more than 500\,000 galaxies in the SDSS DR4\footnote{\tt
  http://www.mpa-garching.mpg.de/SDSS/} \citep{tremonti04}. SDSS
galaxy spectra that have signatures of WR stars span the full range of
emission line ratios measured for other SDSS galaxies
\citep{brinchmann08}. Clearly not all WR regions in field galaxies are
potential progenitor regions for GRBs. Most of these SDSS galaxies are
relatively bright, massive, and hence metal rich; a more appropriate
comparison sample might therefore be a metal poor sample of
galaxies. In Figs.~\ref{fig:n2} and \ref{fig:o2}, the small individual
dots represent galaxies with metallicities between 0.03 and 0.7 solar
from the SDSS \citep{izotov06}. The selection criteria for the various
samples were completely different: the SDSS galaxies were a
flux-limited sample, the metal poor galaxies were a small sub-sample
of the SDSS galaxies with a clear detection of the
[\ion{O}{iii}]~$\lambda$4363 line, while the GRB hosts were not
selected by either of these selection criteria. The GRB hosts occupy
similar regions of the diagram as the metal poor SDSS galaxies,
according to the models, they do not occupy the very metal poor region
of the plots.

Most of the data for the individual \ion{H}{ii} regions in the GRB
980425 host occupy a slightly sub solar metallicity region compared to
the SDSS galaxies. The scatter in the positions is real and reflects
both the age and metallicity spread from one region to the next. As
determined in Sect.~\ref{sect:metal}, the host galaxy has an oxygen
abundance of 0.3--0.8 solar, which is consistent with the model curves
in the left panel, where the small squares occupy the model space
between these two metallicities.

We can compare these model ages for the individual regions in the GRB
980425 host with those determined from the \halpha\ and \hbeta\ EW
measurements in Sect.~\ref{sect:ew}.  The WR region has an age of 3
Myr according to the EW, while younger ages are suggested by the
models in Fig.~\ref{fig:n2}.  This is surprising since a very young
region ($<1$~Myr) is expected to have a high intrinsic extinction,
however the IR emission seen in the Spitzer images \citep{lefloch06}
does not indicate an optically hidden population of young stars that
is not inferred from the \halpha\ emission.  The observations of WR
features imply that stellar populations with ages around 3~Myr must be
present.  The SN region and the integrated spectrum of the galaxy lie
between the 1 and 2 Myr evolutionary model line in the plot,
i.e. similar to the ages determined from the population model fits,
and comparable to the ages derived from the EWs.

%--------------------------------------------
\section{Discussion and conclusions}
\label{sect:disc}
\subsection{Implication for high-z GRB hosts} 
In this paper, we have studied the properties of resolved regions in
the very nearby host galaxy of GRB~980425/SN~1998bw. We found that the
\ion{H}{ii} region where the GRB/SN occurred has the second lowest
oxygen abundance, while the lowest value arises in the WR
region. Apart from the metallicity, other physical properties in the
SN region are representative of the integrated spectrum of the host
galaxy.

The variations, especially those of the metallicity and extinction,
over the face of the galaxy have important implications for the
interpretation of the observations of higher redshift GRB hosts. In
the case that $z>1$ GRB hosts do not contain just one (giant) star
forming region or star cluster, some properties will not be
recoverable when the only information available is the integrated
properties of the entire galaxy. In several GRB hosts, the environment
appears to be complex with a few or several individual star-forming
regions that can only be resolved at the spatial resolution of the HST
\citep[e.g.][]{hjorth02}.

Based on the study of the resolved population of the GRB 980425 host,
we identified the following effects that would occur if the host was
at higher redshift and unresolved

\begin{itemize}
\item The integrated spectra from the separate regions would decrease
  the emission line equivalent widths corresponding to individual
  regions, because older stellar populations would be covered. This
  would then substantially decrease the inferred age of unresolved
  star burst regions.\\

\item The specific star formation rates would be much higher in
  individual \ion{H}{ii} regions than the integrated one where less
  active regions are also covered. \\

\item For the GRB 1998bw host galaxy, the scatter in oxygen abundances
  between different regions would be relatively small compared to the
  overall uncertainty in the calibrations. Abundances would be similar
  to within the 3$\sigma$ uncertainties. For high redshift GRB hosts,
  the integrated spectra may produce a fairly good representation of
  the abundances in the GRB regions.\\

\item We found that the luminosity SSFR introduced in
  \citet{christensen04b} for GRB hosts is a good proxy for the mass
  SSFR, which is more frequently used in the literature. Furthermore,
  the integrated SSFR of the GRB~980425 host is similar to other GRB
  hosts.\\

\item The extinction for small isolated regions can be severely
  underestimated. The overall host has a low reddening of \ebv=0.3
  mag, which is normal for (unresolved) higher redshift GRB hosts. The
  SN region itself has a higher than average value. Relative to
  spectral energy distribution fitting of GRB host broad band data,
  the emission line ratios indicated a higher extinction.\\

\item The emission line ratios for the different regions in the host
  indicated that most other GRB hosts (at $z<0.4$) have properties
  that would seem similar to the WR region in the GRB~980425
  host. This implies that even though we cannot resolve spatially some
  GRB hosts, they must be dominated by very young stars, even though
  we cannot always distinguish any WR features in their spectra.\\

\item The separation of 800 pc between the WR and SN region
  corresponds at $z=1$ to a single pixel with the HST spatial
  resolution of 0\farcs1. Had the host been at $z=1$, we would have
  been able to resolve only two distinct regions: the centre of the
  galaxy and the WR region. In a broad band HST image, the GRB
  location would coincide with the brightest pixel as seen for many
  cosmological GRBs \citep{fruchter06}.

\end{itemize}

\subsection{GRB host metallicities}
If low metallicities are a necessary prerequisite to creating GRBs,
this effect should be immediately apparent from the properties of the
host galaxies. Theoretical models suggest that GRBs correspond to star
forming galaxies with lower than 10\% solar metallicity
\citep{yoon06}.  Models for winds from WR stars are based on Fe as
drivers \citep{vink05}, and likewise the GRB progenitor models are
based on the Fe abundance \citep{yoon05} where no GRB progenitors are
expected from stars with metallicities larger than 0.1--0.3 solar
\citep{woosley06,langer06}.  However, while the models are based on
the Fe metallicity, the emission lines from the host galaxies trace
the oxygen abundance.

As demonstrated in Fig.~\ref{fig:n2} and \ref{fig:o2}, regions with
higher metallicities than the model predictions are generally found
among GRB hosts, although these are still below the solar value
\citep[see also][]{savaglio08}.  A potential bias comes from the fact
that the hosts are generally very faint, so the metallicity
determination has to rely on strong emission line diagnostics, which
have large uncertainties.  The use of strong emission lines to derive
abundances can in some cases infer a higher metallicity estimate than
direct measurement from the [\ion{O}{iii}]~$\lambda$4363 or the
[\ion{N}{ii}]~$\lambda$5755 emission lines. In the WR region, the
strong line diagnostics infer a 0.3 solar oxygen abundance, while the
direct measurement indicates a near solar value. In contrast, the O3N2
diagnostics for the GRB~060218 host indicates a 30\% solar oxygen
abundance, while the direct method gives just 8\% solar
\citep{wiersema07}.

GRB hosts at $z>2$ have metallicities of about 10\% solar in general
as estimated from absorption lines in the afterglows
\citep{savaglio06}.  However, as shown by \citet{vreeswijk07} for the
case of GRB~060418, the rest frame UV absorption lines do not arise in
the immediate GRB progenitor environment, but instead the interstellar
medium of the host galaxy along one single line-of-sight.  On the
other hand, as we have discussed in this paper, the GRB hosts at low
redshift are not necessarily very metal poor. The question is how to
relate the low and high redshift sample of GRB hosts to each other
when the observing techniques are completely different.  Different
biases are included from the assumption that the metallicity of the
progenitor can be constrained from absorption lines in the GRB
afterglow, and that emission lines from the host galaxy can infer the
abundance at the progenitor site. Any of these metallicities may not
be representative of the GRB progenitor itself.  To understand how
these two methods compare with each other, it is necessary to detect
the emission lines and derive abundances from GRB hosts at $z>2$ in a
similar way as performed routinely for local galaxies, as shown in
this paper.  By comparing the absorption metallicity, determined from
the GRB afterglow, with abundances derived from emission lines it can
be verified if the metallicity is representative of the host as a
whole, and to place the GRB hosts in relation to star-forming field
galaxies at high and low redshifts.  The comparison of absorption
lines and emission line spectroscopy for a high redshift GRB host
galaxy can provide a measure of the Fe/O ratio.  Since the GRB hosts
are young and show intense SSFRs, it is likely that the oxygen
abundance is higher than that derived from Fe due to the relatively
small contribution from SN Type Ia. Hence, the determination of the
Fe/O ratio can provide constraints on the star formation history in
galaxies fainter than those generally studied in flux-limited surveys.

\begin{acknowledgements}
LC acknowledges the hospitality of the people at the Dark Cosmology
Center in Copenhagen, where much of the data analysis was done. The
Dark Cosmology Centre is funded by the Danish National Research
Foundation.  PMV acknowledges the support of the EU under a Marie
Curie Intra-European Fellowship, contract MEIF-CT-2006-041363. JS is a
Royal Swedish Academy of Sciences Research Fellow supported by a grant
from the Knut and Alice Wallenberg Foundation. Support for ELF's work
was provided by NASA through the Spitzer Space Telescope Fellowship
Program. This research has made use of the GHostS database
(www.grbhosts.org), which is partly funded by Spitzer/NASA grant RSA
Agreement No. 1287913. We thank G\"oran \"Ostlin for the \halpha\ data
check, Sandra Savaglio for useful suggestions and discussions, and
Damien Le Borgne for sharing the Z-Peg code, and finally Avishay
Gal-Yam for sharing the Gemini spectrum of the GRB~060614 host. We
thank the referee for a constructive report.
\end{acknowledgements}

\bibliography{paper}
\end{document}